\documentclass[11pt,journal]{IEEEtran}
\usepackage{amsmath,amsfonts,amssymb,mathtools}
\usepackage{algorithmic}
\usepackage{algorithm}
\usepackage{array}
\usepackage[caption=false,font=footnotesize]{subfig}
\usepackage{textcomp}
\usepackage{stfloats}
\usepackage{url}
\usepackage{verbatim}
\usepackage{graphicx}
\usepackage{tikz}
\usepackage{cite}
\usepackage{wrapfig}
\usepackage[table]{xcolor}

\usepackage{booktabs}
\usepackage{tabularx}
\usepackage{amssymb}

\DeclareMathOperator{\diag}{diag}

\DeclareMathOperator{\IID}{i.i.d.}

\DeclareMathOperator*{\argmin}{arg\,min}

\newcommand{\p}[1]{p_{X}}
\newcommand{\q}[1]{h^{#1}}
\newcommand{\Q}[1]{Q_{\hat{X}|X}^{#1}}
\newcommand{\Qm}[1]{q_{\hat{X}}^{#1}}
\newcommand{\va}[1]{
    \ifthenelse{\isempty{#1}} 
        {v} 
        {v^{#1}}
}
\newcommand{\tq}[1]{\hat{h}^{#1}}
\newcommand{\tQ}[1]{\hat{Q}_{\hat{X}|X}^{#1}}
\newcommand{\tQm}[1]{\hat{q}_{\hat{X}}^{#1}}

\newcommand{\df}{\partial f}

\newcommand{\setSym}{\mathcal{X}}

\newcommand{\A}[2][]{
    \ifthenelse{\isempty{#1}} 
        {A^{#2}(x,y,s)} 
        {A^{#2}(x,#1,s)}
}

\newcommand{\tA}[2][]{
    \ifthenelse{\isempty{#1}} 
        {\hat{A}^{#2}(x,y,s)} 
        {\hat{A}^{#2}(x,#1,s)}
}

\newcommand{\bA}[2][]{
    \ifthenelse{\isempty{#1}} 
        {A^{#2}[u](x,y,s)} 
        {A^{#2}[u](x,#1,s)}
}

\newcommand{\ssum}[1]{ 
    \sum_{#1 \in \setSym}
}

\newcommand{\ssumh}[1]{ 
    \sum_{#1 \in \setSym}
}

\newcommand{\cx}[1][]{
    \ifthenelse{\isempty{#1}} 
        {c(y)} 
        {c(#1)}
}

\newtheorem{theorem}{\bfseries Theorem}
\newtheorem{lemma}{\bfseries Lemma}
\newtheorem{definition}{\bfseries Definition}
\newtheorem{corollary}{\bfseries Corollary}

\newtheorem{remark}{\bfseries Remark}

\DeclareMathOperator{\KL}{D_{KL}}

\DeclareMathOperator{\WTS}{W_2^2}

\DeclareMathOperator{\ND}{\mathcal{N}}

\newcommand{\RG}{R^G}
\newcommand{\RGi}{R^G_i}

\newcommand{\eigvi}[1]{{\lambda_{#1,i}}}
\newcommand{\eigviSq}[1]{{\lambda^2_{#1,i}}}
\newcommand{\mean}[1][]{ \ifthenelse{\isempty{#1}}{\mu}{\mu_{#1}}}
\newcommand{\var}[1][]{\ifthenelse{\isempty{#1}}{\sigma^2} {\sigma^2_{#1}}}
\newcommand{\Cov}[1][]{\ifthenelse{\isempty{#1}}{\Sigma} {\Sigma_{#1}}}

\newcommand{\stv}[1][]{\ifthenelse{\isempty{#1}}{\sigma}{\sigma_{#1}}}
\newcommand{\stveigvi}[1][]{\ifthenelse{\isempty{#1}}{\sqrt{\eigvi{\Cov[X]}}}{\sqrt{\eigvi{\Cov[#1]}}}}
\newcommand{\lamb}[1][]{\ifthenelse{\isempty{#1}}{\mathcal{W}} {\mathcal{W}_{#1}}}
\newcommand{\hX}{\hat{X}}

\newcommand{\pdf}[1]{ p_{#1} }

\newcommand{\vecD}{\mathbf{D}}
\newcommand{\vecP}{\mathbf{P}}

\newcommand{\E}[2][]{\mathbb{E}_{#1} \left[ #2 \right]}

\begin{document}

\title{Rate-Distortion-Perception Theory: Redefining the Fundamental Limits of Information Representation}

\author{Photios A. Stavrou,~\IEEEmembership{Senior Member,~IEEE,} Giuseppe Serra,~\IEEEmembership{Graduate Student Member,~IEEE,} Marios Kountouris,~\IEEEmembership{Fellow,~IEEE,}
}


\maketitle

\begin{abstract}
 Classical rate–distortion (RD) theory has long established the fundamental limits of lossy compression by quantifying the minimum number of bits required to represent a source under a prescribed distortion constraint. However, widely used distortion measures such as mean-squared error often fail to capture perceptual quality or semantic validity, which are increasingly central in modern learning-driven applications. Rate–distortion–perception (RDP) theory extends the RD framework by introducing perception as a third fundamental axis, quantified via distributional similarity between the source and reconstructed signals, leading to the rate–distortion–perception function (RDPF). This tutorial provides a structured overview of the coding principles underlying perception-aware lossy compression and surveys recent achievability results under different randomness assumptions. It then presents a unifying optimization viewpoint for computing the RDPF as defined by Blau and Michaeli, for both discrete and continuous sources under broad families of perceptual constraints, including $f$-divergences, $\alpha$-divergences, and Wasserstein-based metrics. Special attention is given to computational tools such as alternating minimization schemes, Newton-based methods, and convex optimization formulations, as well as to analytically tractable cases such as Gaussian sources and the perfect-realism regime. {Unlike recent broad surveys that emphasize generative architectures and AI-empowered communication systems, this tutorial focuses on the coding-theoretic and computational machinery needed to characterize, compute, and interpret the RDP limits}. Finally, the tutorial outlines promising research directions at the intersection of information theory, neural compression, robust source coding, and perception-aware networked control systems.
\end{abstract}


\section{Introduction}\label{sec:introduction}
\IEEEPARstart{C}{lassical} rate–distortion (RD) theory, introduced by Shannon in the 1950s \cite{shannon:59}, has long served as a cornerstone for understanding the trade-off between compression efficiency and reconstruction fidelity. In this theory, the rate–distortion function (RDF) specifies the minimum number of bits required to represent a source subject to a prespecified average distortion constraint. For decades, this framework has guided the design of source coding systems and has been instrumental in shaping the development of digital communications and signal compression. However, in practice, commonly used distortion measures, such as mean-squared error (MSE) or Hamming distance, often fail to reflect how humans or intelligent agents perceive quality, utility, and realism. Reconstructions with low distortion may still appear perceptually implausible or semantically misleading, whereas reconstructions with slightly higher distortion may be subjectively preferable or more useful for a given task.

Rate–distortion–perception (RDP) theory was recently introduced in \cite{blau:2019} to bridge this gap. By adding perception as a third fundamental dimension to the trade-off, RDP theory extends the classical RDF formulation to account not only for bit rate and distortion, but also for the perceptual fidelity of reconstructions. Perceptual fidelity is typically quantified using measures of distributional similarity between reconstructed signals and the natural statistics of the source data, helping ensure that reconstructions are not only accurate but also perceptually plausible and semantically meaningful. This extension gives rise to the rate–distortion–perception function (RDPF), which characterizes the fundamental limits achievable when all three objectives, i.e., rate, distortion, and perception, are jointly optimized. 

\begin{figure}[t]
        \centering
\includegraphics[width = \linewidth]{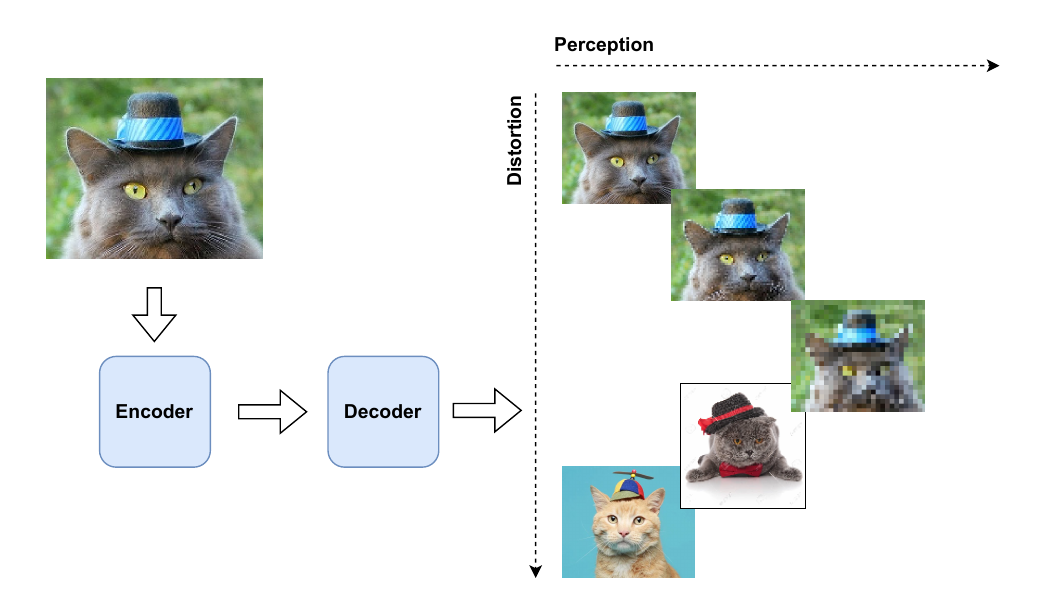}
\caption{When applying rate–distortion–perception (RDP) theory to image compression, the goal is not necessarily to obtain the most pixel-accurate reconstruction. Instead, the objective is to preserve the perceptually important characteristics of the image. For example, even in a high-distortion regime, the reconstructed image may not appear “highly distorted” to a human observer; rather, it can still convey the essential semantic content and recognizable features, such as clearly depicting a cat wearing a hat.} \label{fig:0}
\end{figure}

The timeliness of this topic is underscored by the growing importance of AI-driven information processing and semantic communications \cite{kountouris:2021,strinati:2024}. In modern applications, such as image and video compression with generative models, speech synthesis, immersive media transmission, and reinforcement learning–based control, perceptual realism and semantic coherence are increasingly prioritized over strict numerical fidelity. For example, deep generative codecs often produce reconstructions that deviate from the original pixel values yet remain subjectively indistinguishable to human observers and semantically correct for downstream tasks. Similarly, semantic communication frameworks aim to convey the most important and relevant information rather than exact signal replicas, aligning naturally with the principles of the RDP paradigm (see Fig.~\ref{fig:0}).

Beyond communication and compression, RDP theory also has direct implications for networked control systems \cite{hespanha:2007}. In feedback loops where control decisions rely on compressed or transmitted measurements, it is often insufficient for reconstructions to minimize distortion alone; they must also preserve a perceptually plausible representation of the system state to help guarantee stability and safe operation. This requirement is particularly critical in autonomous systems, robotics, and cyber-physical systems, where intelligent controllers interact with uncertain environments using compressed or partial sensory data. The RDP framework offers a principled way to quantify the minimal communication resources needed to guarantee not only stability and accuracy, but also task-aware and perceptually meaningful performance.

In addition, the growing convergence of machine learning, optimization, and information theory further underscores the relevance of RDP theory. Contemporary compression and communication systems are increasingly built on neural networks \cite{lei:2025}, adversarial training, and distribution-matching objectives, all of which relate directly to the perceptual axis formalized by the RDPF. The ability to rigorously characterize and compute the trade-offs among bit rate, distortion, and perceptual fidelity, therefore, provides a solid theoretical foundation for analyzing and benchmarking these emerging technologies.

In this tutorial, we first provide an overview of existing works that propose coding-theorem constructions leading to various RDPF characterizations under different assumptions (see Section~\ref{sec:coding_principles}). Our main focus is to compile a collection of generic optimization and computational tools for computing the original RDPF characterization introduced in \cite{blau:2019}, for a variety of independent and identically distributed ($\IID$) source models and divergence measures. While the methods presented here are not exhaustive for every possible setting, their generality makes them applicable to a broad range of extensions and more challenging problems that lie beyond the scope of this tutorial (see Section~\ref{sec:computational_aspects}). We conclude with our perspective on several open questions motivated by our ongoing research on RDP theory (see Section~\ref{sec:future_work}).

{
\paragraph*{Relation to existing survey literature} Existing survey-style treatments of RDP have primarily emphasized either the broader role of perception constraints in generative compression and intelligent communications \cite{niu:2025} or the position of RDP theory within the wider landscape of AI-era compression \cite{chen:2025}. In contrast, the present tutorial focuses specifically on the coding principles and computational machinery underlying the RDPF.}


\section{Coding Principles Behind Perception-Aware Lossy Compression}\label{sec:coding_principles}

In this section, we review existing coding theorems that demonstrate the achievability of the information-theoretic measure introduced in \cite{blau:2019}. For completeness, we also discuss the role of common randomness in coding techniques, noting that, in many cases, common randomness yields achievability results that do not coincide with the characterization in \cite{blau:2019}.

We then present the first information-theoretic characterization of the RDP trade-offs for general alphabets. This characterization is the main focus of this tutorial.

\begin{definition} \cite{blau:2019} (RDPF) \label{def:rdpf}
Let a source $X$ be a random variable on a measurable space $(\mathcal{X}, \mathbb{B}(\mathcal{X}))$ distributed according to the probability measure $\pdf{X} \in \mathcal{P}(\mathcal{X})$.
Then, the RDPF for a source {$X\sim{p}_X$} under the distortion measure $\Delta:\mathcal{X}^2 \to \mathbb{R}^+_0$ and divergence function $d: \mathcal{P}(\mathcal{X}) \times \mathcal{P}(\mathcal{X}) \to  \mathbb{R}^+_0$ is defined as follows:
\begin{align}
        R(D,P)\triangleq & \min_{\substack{Q_{\hat{X}|X}\\ \E{\Delta(X,\hX)} \le D\\ D(p_X||p_{\hX}) \le P}}  I(X,\hX) \label{eq.sec2:rdpf}  
\end{align}
where $D\geq{0}$, $P\geq{0}$, and the minimization is over all conditional distributions $Q_{\hX|X}: \mathcal{X} \to \mathcal{P}(\mathcal{\hat{X}})$.
\end{definition}
\par Evidently, the RDPF generalizes the classical RDF by providing, in addition to the minimum number of bits per sample, a measure of how ``natural'' or ``perceptually plausible'' the reconstruction appears. 
Interestingly, the RDPF shares many of the same functional properties as the classical RDF. The only notable difference lies in the divergence-based perception fidelity term, which, under the general assumption that the divergence is convex in its second argument \cite{blau:2019}, still yields a convex optimization problem in \eqref{eq.sec2:rdpf}.

\subsection{Coding theorems resulting in \eqref{eq.sec2:rdpf}}\label{sec2:subsec:1} 

Recent years have seen significant progress in understanding when the information measure in \eqref{eq.sec2:rdpf}, as well as the special case of a perfect perception constraint, can actually be achieved by a coding scheme. Early works in \cite{matsumoto:2018,matsumoto:2019} approached the problem using information-spectrum methods \cite{han:2003}, which provide powerful tools for characterizing RDP tradeoffs without assuming memory or stationarity. Around the same time, \cite{theis:2021} used the strong functional representation lemma, together with the idealized assumption of infinite common randomness \cite{li:2018}, to show that \eqref{eq.sec2:rdpf} is asymptotically achievable using stochastic, variable-length codes. More recently, \cite{chen:2022} revisited the role of randomness in the coding process and distinguished among common randomness, private randomness, and purely deterministic encoding. This perspective clarifies how different types of randomness affect the operational rates of achievable schemes, leading to weaker sufficient conditions than those in \cite{theis:2021}. Under these relaxed conditions, \eqref{eq.sec2:rdpf} remains asymptotically achievable for $\IID$ sources, even when the target perception constraint $P$ and distortion constraint $D$ are not freely chosen.

\subsection{Beyond infinite common randomness}\label{sec2:subsec:2}

While mathematically convenient, especially in random binning, channel simulation, and perception-aware compression, one may argue that the assumption of infinite common randomness is rather idealized, as real systems can only coordinate through finite seeds or explicitly transmitted randomness. By contrast, constraining the system to finite common randomness yields performance guarantees that more accurately reflect practical implementations. It forces the encoder to account for the randomness needed to reproduce perceptual variability, thereby revealing the true tradeoff between rate, distortion, and perceptual fidelity. Moreover, limited common randomness aligns with modern GAN-based generative compression systems, which rely on small shared seeds or deterministic decoders rather than idealized infinite-randomness sources, further motivating the need to study achievability under more realistic randomness constraints. 
\par Recent work has therefore begun to investigate the fundamental limits of perception-aware compression with limited common randomness. A foundational contribution is \cite{saldi:2015}, which considered output-constrained lossy source coding with limited common randomness. This framework differs from classical lossy source coding in that the reconstructed symbols must be $\IID$ with a prescribed marginal distribution, thereby imposing a form of perceptual realism at the distributional level. Building on \cite{saldi:2015}, the work of \cite{wagner:2022} investigated the role of limited common randomness in the perfect realism case, whereas the recent work of \cite{xie:2025} studied perception-aware compression under a squared-error distortion measure and perception constraints based on the Kullback–Leibler (KL) divergence and the squared Wasserstein-2 distance. Notably, some special cases that assume jointly Gaussian sources reveal closed-form expressions different from those obtained by solving \eqref{eq.sec2:rdpf}. In a companion paper, \cite{xie:2026} specialized their analysis to Gaussian sources, deriving bounds on the operational rates and exploring RDP tradeoffs using entropy-constrained scalar quantization. Finally, the work of \cite{qu:2025tit} studies the RDP tradeoffs in the Wasserstein space, under various cases of common randomness and perception constraints.
\par 

\section{Optimization Viewpoints and Computational Aspects of \eqref{eq.sec2:rdpf}}\label{sec:computational_aspects}

\par In this section, we present general-purpose optimization and computational tools for solving \eqref{eq.sec2:rdpf} and its special case of perfect realism, for both discrete and continuous $\IID$ sources under single-letter distortion constraints and a wide range of perception constraints. Although our focus is on computing \eqref{eq.sec2:rdpf}, the methodologies presented here extend well beyond this class of problems. For example, with appropriate modifications to the objective function and feasible set, our optimization tools can be applied to compute the distortion-rate-perception function under finite common randomness, either optimally or near-optimally. Moreover, they can be used to derive bounds on the RDPF when the perception constraint is not tensorizable.

\subsection{Discrete Sources subject to $f$-divergence perceptual constraints}\label{section:iiia}

We start by providing a tutorial on the computation of \eqref{eq.sec2:rdpf}, when the distortion is single-letter and the perception constraint belongs to the family of $f-$divergences.

\paragraph*{Statistical divergences and the family of f-divergences} Statistical divergences are fundamental measures used in information theory and statistics to quantify the dissimilarity between probability distributions. In their general definition, a divergence on $\mathcal{P}({\cal X})$ is a function $D: \mathcal{P}({\cal X})^2 \to \mathbb{R}_{0}^+$ such that $D(p_X||q_X) \ge 0$ for all $p_X,q_X \in \mathcal{P}({\cal X})$, with equality if and only if $p_X = q_X$. The family of $f$-divergences was first introduced in \cite{renyi:1961} (see also \cite{fdiv-csizar}). This rich class of divergences is defined as follows.

\begin{definition}\cite{fdiv-csizar} {($f$-divergence)}
    Let $f:(0,\infty) \to \mathbb{R}$ be a convex function with $f(1) = 0$. The $f$-divergence $D_f(\cdot||\cdot)$ associated with $f$ is defined as
    \begin{align*}
        D_f(p_X||q_X) \triangleq \sum_{x\in{\cal X}} q_Xf\left(\frac{p_X}{q_X}\right),~p_X,q_X \in \mathcal{P}({\cal X})
    \end{align*}
    under the conventions
    \begin{align*}
        \text{(i) } f(0) &= \lim_{ x \to 0^+} f(0),  \qquad \text{(ii) } 0f\left( \frac{0}{0} \right) = 0, 
        \\ &\text{(iii) } \forall a \ge 0,~0f\left( \frac{a}{0} \right) = af'(\infty).  
    \end{align*}

\end{definition}
Indicative examples of commonly used divergence functions that belong to the class of $f$-divergences include
\begin{itemize}
    \item {\bf KL divergence $D_{KL}(\cdot||\cdot)$}, obtained by setting $f(x) = x\log(x)$,
        \begin{align*}
            D_{KL}(p_X||q_X) = \sum_{x\in{\cal X}} p_X\log\left(\frac{p_X}{q_X}\right)
        \end{align*}
    \item {\bf Jensen-Shannon divergence $D_{JS}(\cdot||\cdot)$}, obtained by setting $f = x\log \left(\frac{2x}{x+1} \right) + \log \left(\frac{2}{x+1} \right)$,
        \begin{align*}
            D_{JS}(p_X||q_X) = & D_{KL}\left(p_X \Big|\Big|\frac{p_X+q_X}{2}\right) \nonumber\\
            &+\qquad D_{KL}\left(q_X\Big|\Big|\frac{p_X+q_X}{2}\right)
        \end{align*}
    \item  {\bf Total Variation (TV) $TV(\cdot||\cdot)$}, obtained by setting $f = \frac{1}{2}|x - 1|$,
        \begin{align*}
            TV(p_X||q_X) =  \frac{1}{2} \sum_{x\in{\cal X}} |p_X - q_X|
        \end{align*}
     \item {\bf $\alpha$-divergence $D_{\alpha}(\cdot||\cdot)$}, where $f$ is parameterized by $\alpha \in \mathbb{R}$, denoted by $f_\alpha$, such that
        \begin{align}
            D_{\alpha}(&p_X||q_X) = \sum_{x\in{\cal X}} q_Xf_{\alpha}\left(\frac{p_X}{q_X}\right)\label{cziscar_a_divergence}
            \end{align}
where            
     \begin{align*}
     f_{\alpha}(x) &= \
            \begin{cases}
                \frac{x^\alpha - \alpha x - (1-\alpha)}{\alpha(\alpha -1)} \qquad \text{if } \alpha \neq 0, \alpha \neq 1 \\ 
                x\ln(x) - x + 1 \qquad \text{if } \alpha = 1\\
                -\ln(x) + x - 1 \qquad \text{if } \alpha = 0\\
            \end{cases}.
        \end{align*}
\end{itemize}
The class of $f$-divergences possesses a number of attractive functional properties, most notably joint convexity and invariance, which enhance their usefulness. A comprehensive mathematical discussion can be found in \cite{sason:2018}.

\paragraph*{Alternating Minimization} \label{sec: AM and BA-type algorithms} The alternating minimization method is an optimization framework for minimizing functions of two variables subject to constraints. Consider the following optimization problem
\begin{align*}
    \min_{\substack{ x \in {\cal X}, y \in {\cal Y} }}  f(x,y)
\end{align*}
where ${\cal X}$ and ${\cal Y}$ are two arbitrary non-empty sets and the function $f(x,y)$ satisfies $f(x,y)\in(-\infty, +\infty]$ for each $x\in{\cal X}$ and $y\in{\cal Y}$. Furthermore, we assume that, for each $x\in{\cal X}$, there exists $y\in{\cal Y}$ with $f(x,y)<+\infty$, implying that $s\triangleq \inf_{\substack{ x \in {\cal X}, y \in {\cal Y} }} f(x,y) < +\infty$.
Depending on the setting, one may also assume the existence or uniqueness of a minimizer $(x^*,y^*)$ such that $f(x^*,y^*) = s$. 

The goal of the alternating minimization method is to construct a sequence $\{(x^{(n)},y^{(n)})\}$ such that $\displaystyle \lim_{n \to \infty} f(x^{(n)},y^{(n)}) = s$. Under specific conditions, such a sequence can be defined using the solutions of two subproblems: for $x_i \in {\cal X}$, $h(x_i) = \displaystyle \argmin_{y \in {\cal Y}}  f(x_i,y)$, and for $y_i \in {\cal Y}$, $g(y_i) = \displaystyle \argmin_{x \in {\cal X}}  f(x,y_i)$. Starting from an initial point $y^{(0)}$, we can define the $n$-{th} element of the sequence as
\begin{align*}
    x^{(n)} = g(y^{(n-1)}) \qquad y^{(n)} = h(x^{(n)}) \quad \text{for} ~n = 1,2,\ldots.
\end{align*}
Depending on the problem, various sufficient conditions for the existence of the sequence limit and its optimality have been studied; see, for instance, \cite{csiszar:1984,yeung:2008,grippo:2000}. Early applications of alternating minimization to information-theoretic optimization problems can be traced to the foundational contributions of \cite{arimoto:1972, blahut:72}.

In what follows, we state some theoretical results related to the computation of \eqref{eq.sec2:rdpf} for discrete $\IID$ sources. For clarity, we use square brackets to indicate functional dependence between mathematical objects; e.g., $p[h]$ and $Q[h]$ denote the functional dependence of a distribution $p \in \mathcal{P}({\cal X})$ or a transition matrix $Q \in \mathcal{Q}({\cal X})$ on another distribution $h \in \mathcal{P}({\cal X})$.

The first result provides a parametric characterization of the solution to \eqref{eq.sec2:rdpf} as a double minimization problem.
\begin{lemma} \cite{serra:2025tit} (Double minimization)\label{lemma:DoubleMinimization}
Let $D \ge 0$ and $P \ge 0$, and let $D(\cdot||\cdot)=D_f(\cdot||\cdot)$. Moreover, let $s=(s_D, s_P)$ with $s_D\geq{0}$ and $s_P\geq{0}$ denoting the Lagrangian multipliers associated with the distortion and perception constraints in \eqref{eq.sec2:rdpf}. Then \eqref{eq.sec2:rdpf} can be cast as the following double minimization: 
\begin{align}
\begin{split}
R(D,P) =  &\min_{\substack{\Q{} \in \mathcal{Q}(\setSym) \\ \q{} \in \mathcal{P}(\setSym)} } D_{KL}(\p{} \Q{} || \p{} \q{} ) \\ \quad&+ s_D\left(\E[]{\Delta(X,\hat{X})} - D \right)\\
&+ s_P( D_f(\p{}||q_{\hat{X}}) - P)
\end{split}\label{eq: double_min}
\end{align}
where  $D =\E[\Q{*}]{\Delta(X,\hat{X})}$, $P=D_f(\p{}||\q{*})$, with $(\Q{*},\q{*})$ being the pair achieving the minimum and $q_{\hat{X}}$ denotes the reconstruction distribution induced by $\hat{X}$.\\ 
Furthermore, for fixed $\Q{}$, the right-hand side (RHS) of \eqref{eq: double_min} is minimized by
\begin{align}
\q{}[\Q{}](\hat{x}) = \sum_{x \in \setSym}  \p{}(x) \Q{}(\hat{x}|x) \label{eq: optimization qx}
\end{align}
whereas, for fixed $\q{}$, the RHS of \eqref{eq: double_min} is minimized by
\begin{align}
\Q{}[\q{}](\hat{x}|x) &= \frac{\q{}(\hat{x}) \cdot A[\Qm{}[\q{}]](x,\hat{x},s)}{\sum_{i \in \setSym} \q{}(i) \cdot A[\Qm{}[\q{}]](x,i,s)} \label{eq: ParametricQ}
\end{align}
in which
\begin{align}
A[u](x,\hat{x},s) &=e^{-s_D \Delta(x,\hat{x}) - s_P g(\p{}(\hat{x}), u(\hat{x}))} \label{eq: optimization Qxx - A} \\ 
\Qm{}[u](\hat{x}) &= \ssum{x} \Q{}[u](\hat{x}|x) \p{}(x) \label{eq: MarginalQ}\\
g(x,\hat{x}) &=  f\left(\frac{x}{\hat{x}}\right) - \frac{x}{\hat{x}} \df\left(\frac{x}{\hat{x}}\right).\nonumber 
\end{align} 
\end{lemma}

Building on the results of Lemma \ref{lemma:DoubleMinimization}, we now construct an alternating minimization procedure, henceforth referred to as optimal alternating minimization (OAM), and show that it converges to a point on the RDPF $R(D,P)$ as defined in \eqref{eq.sec2:rdpf}. 

\begin{theorem}\cite{serra:2025tit} (OAM) \label{th:OAM}
Let the Lagrangian multipliers $s=(s_D, s_P)$ with $s_D \ge 0$, $s_P \ge 0$  be given. Let $\q{(0)}$ be any probability vector with strictly positive components and let $\Q{(n+1)} \equiv\Q{}[\q{(n)}]$ and $\q{(n+1)} \equiv\Qm{}[\q{(n)}]$ be functions of the current iteration $\q{(n)}$ as defined in \eqref{eq: ParametricQ} and \eqref{eq: MarginalQ}, respectively.
Then, as $n\xrightarrow{}\infty$, we have
\begin{align*}
&D(\Q{(n)})\xrightarrow{} D_s,~P(\Q{(n)})\xrightarrow{} P_s\nonumber\\
&\qquad I(\p{}, \Q{(n)}) \xrightarrow{} R(D_s,P_s).
\end{align*}
\end{theorem}

Unfortunately, the OAM scheme does not readily admit a Blahut-Arimoto-type algorithmic implementation \cite{blahut:72}. The reason lies in the fact that the parametric dependencies underlying \eqref{eq: ParametricQ} and \eqref{eq: MarginalQ} induce an implicit dependency of $\q{(n+1)}$ on itself. In particular, one obtains
    \begin{align}
        \frac{\q{(n+1)}(\hat{x})}{\q{(n)}(\hat{x})} &=\sum_{x \in \setSym} \frac{ \p{}(x) e^{-s_D \Delta(x,\hat{x}) - s_P g\left( \p{}, {\q{(n+1)}}, \hat{x} \right) }}{\sum_{i\in\hat{\cal X}} \q{(n)}(i) e^{-s_D \Delta(x,i) - s_P g\left( \p{}, {\q{(n+1)}}, i\right)} } \nonumber\\ 
        &= c[\q{(n)},\q{(n+1)}](y),\label{eq:rec_unfold}
    \end{align}
which makes the update implicit.
As a result, we explore alternative methods for computing the alternating minimization iterations.

\noindent{\bf Method 1: Newton-based Alternating Minimization (NAM)}  We first demonstrate that the update for $\q{(n+1)}$, namely \eqref{eq:rec_unfold}, can be cast as a root finding problem.

\begin{lemma} \cite{serra:2025tit}\label{lemma: Newton-Function Definition}
Let $\q{(n+1)}$ be defined as in Theorem \ref{th:OAM}, and let $T:\mathbb{R}^{|\setSym|} \to \mathbb{R}^{|\setSym|}$ be the vector-valued function defined by
\begin{align}
    T[\q{(n)}, u](i) \triangleq u(i) - \q{(n)}(i) \cdot c[\q{(n)},u](i),~ \forall i \in \setSym \label{eq: NewtonRootFunctional}
\end{align} 
where $c[\cdot, \cdot]$ is given by
\begin{align}
 c[u, r](y) =\frac{}{} \ssum{x} \frac{ \p{}(x) A[r](x,y,s)}{\ssumh{i} u(i) A[r](x,i,s)}. \label{eq: c_coeff}
\end{align}
Then $\q{(n+1)}$ is a root of $T[\q{(n)}, \cdot]$, i.e., $T[\q{(n)},\q{(n+1)}] = 0$.
\end{lemma}

The use of Newton’s method requires both the existence and invertibility of the Jacobian matrix $J_T$ associated with the mapping $T$ \cite[Section 10.2]{burden:2015}.
In our setting, ensuring the existence of $J_T$ imposes a stronger smoothness requirement on the divergence measure.
In particular, the divergence $D_f(\cdot || \cdot)$ must be twice differentiable with respect to its second argument.
Although this assumption limits the full generality of the NAM method, it is not overly restrictive in practice, since most commonly used divergences, including the vast majority of $f$-divergences, satisfy this level of smoothness.

Under this differentiability condition, we can further show that the Jacobian $J_T$ is indeed invertible. This property is established in the following lemma.

\begin{lemma} \cite{serra:2025tit} \label{lemma: Newton-Jacobian}
    Let $T[\q{(n)}, \cdot]$ be the function defined in \eqref{eq: NewtonRootFunctional}, and assume that the divergence measure $D_f(\cdot||\cdot)$ is twice differentiable with respect to its second argument. Then, the Jacobian $J_T: \mathbb{R}^{|\setSym|} \to \mathbb{R}^{|\setSym| \times |\setSym|} $ of the functional $T[\q{(n)}, \cdot]$, defined as $J_T[\q{(n)}, u] \triangleq \left[ \frac{\partial T[\q{(n)}, v](i)}{\partial v(j)} \Big{|}_{u} \right]_{(i,j) \in \setSym^2}$, is positive definite and has the form
    \begin{align}
        J_T[\q{(n)}, u] = I + \left(C[\q{(n)}, u] - M[\q{(n)}, u]\right) \cdot \Gamma[\q{(n)}, u] \label{eq: NewtonFunctionJacobian}
    \end{align}
    where
   \begin{align}
       &M[\q{(n)}, u] = \nonumber\\
       &~\left[ \q{(n)}(i) \ssum{x} \p{}(x) \frac{A[u](x,i,s) \cdot A[u](x,j,s)}{ \left(\ssum{k} \q{(n)}(k) A[k](x,k,s)\right)^2}  \right]_{(i,j) \in \setSym^2} \nonumber
  \end{align}
    \begin{align} 
        \Gamma[\q{(n)}, u] &=  s_P \diag\left[ \q{(n)}(i) \cdot \frac{\partial^2 D_f(\p{}||v)}{\partial v(i)^2}\bigg{|}_{u} \right]_{i \in \setSym} \nonumber\\
        C[\q{(n)}, u] &= \diag\Big[c[\q{(n)}, u](i) \Big]_{i \in \setSym}.\nonumber
    \end{align}
\end{lemma}

Next, we provide the structure of the Newton root-finding iteration applied to the functional $T(\cdot)$, which, as shown in Lemma \ref{lemma: Newton-Function Definition}, yields an approximation of $\q{(n+1)}$.

\begin{theorem} \cite{serra:2025tit} {(Newton's method)} \label{th: Newton Approximation Method}
Assume that the divergence measure $D_f(\cdot||\cdot)$ is twice differentiable with respect to its second argument, and let $\q{(n+1)}$ and $\q{(n)}$ be defined as in Theorem \ref{th:OAM}. Let $T[\q{(n)}, \cdot]$ and $J_T[\q{(n)}, \cdot]$ be as defined in \eqref{eq: NewtonRootFunctional} and \eqref{eq: NewtonFunctionJacobian}, respectively. Furthermore, let the sequence $\{u^{(k)}\}_{k=1,2,\ldots}$  for some initial point $u^{(0)} \in \mathbb{R}^{|\setSym|}$ be defined by
\begin{align*}
u^{(k + 1)} \triangleq u^{(k)} - \left(J_T[\q{(n)}, u^{(k)}] \right)^{-1} \cdot T[\q{(n)}, u^{(k)}].
\end{align*}
Then, $\lim_{k \to \infty} u^{(k)} = \q{(n+1)}$.
\end{theorem}

Applying Theorem~\ref{th: Newton Approximation Method} to the OAM scheme introduced in Theorem~\ref{th:OAM} yields an NAM-based algorithm for computing \eqref{eq.sec2:rdpf} for discrete sources with single-letter distortion measures and perception constraints expressed via an $f$-divergence (see \cite[Algorithm~1]{serra:2025tit}), with exponential convergence in the number of iterations. However, resolving the main technical challenges of the OAM scheme within the NAM framework requires restrictive assumptions on the admissible perception metrics. In the sequel, we therefore present an alternative minimization scheme that circumvents these limitations.

\noindent{\bf Method 2: Relaxed Alternating Minimization (RAM)} An alternative way to address the implementation challenges of the OAM scheme is to adopt a relaxed formulation of the OAM iterations. By introducing an auxiliary design variable $\upsilon$ into \eqref{eq: ParametricQ}, we obtain an approximation of the original OAM scheme, which we term the RAM scheme. The key advantage of the RAM scheme is that, with an appropriate choice of $\upsilon$, the resulting iterative procedure is directly implementable and does not rely on additional continuity assumptions on the perception constraints, while still guaranteeing convergence to a globally optimal solution. The following theorem formally presents the RAM iterative scheme.

\begin{theorem} \cite{serra:2025tit} (RAM) \label{th:RAM}  Let the Lagrange multipliers $s=(s_D, s_P)$ with $s_D \ge 0$, $s_P \in [0, s_{P, \max}]$ be given and define 
\begin{align}
    \tQ{}[u](\hat{x}|x)  & \triangleq \frac{ u(y) A[\upsilon[u]](x,\hat{x},s)}{ \ssumh{i} u(i) A[\upsilon[u]](x,i,s)} \label{eq:tQ}\\
    \tQm{}[u](\hat{x})  & \triangleq \ssum{x} \tQ{}[u](\hat{x}|x) \p{}(x) \label{eq:tQm}
\end{align}
where $A[\cdot]$ is defined in \eqref{eq: optimization Qxx - A} and $v[\cdot]:\mathcal{P}(\setSym) \to \mathcal{P}(\setSym)$ is any functional that maps a probability distribution to a probability distribution.
Let $\tq{(0)}$ be any probability vector with nonzero components and let $\tQ{(n+1)} \equiv \tQ{}[\tq{(n)}]$, $\tq{(n+1)} \equiv \tQm{}[\tq{(n)}]$, and $\upsilon^{(n)} = v[\tq{(n)}]$. Then, as $n\xrightarrow{}\infty$, we obtain
\begin{align*}
&D(\tQ{(n)}) \xrightarrow{} D_s,~P(\tQ{(n)}) \xrightarrow{} P_s\\
&\qquad I(\p{}, \tQ{(n)}) \xrightarrow{} R(D_s,P_s)
\end{align*}
provided that $\lim_{n\rightarrow\infty} || \q{(n+1)} - \upsilon^{(n)}|| = 0$ with at least a linear rate of convergence.
\end{theorem}

Theorem \ref{th:RAM} facilitates the implementation of the alternating minimization scheme by introducing an auxiliary variable $\upsilon[\q{(n)}]$, which serves as an approximation of the true update $\q{(n+1)}$ while remaining a function solely of the current iterate $\q{(n)}$. However, depending on the choice of $\upsilon$, this approximation may impose restrictions on the admissible range of the Lagrange multiplier $s_P$, which can, in turn, affect the convergence guarantees. The algorithmic realization of the RAM approach is presented in \cite[Algorithm~2]{serra:2025tit}, where exponential convergence is established under suitable conditions.

We conclude this section with a technical remark highlighting the differences between Methods 1 and 2.
\begin{remark}[NAM versus RAM]
The main advantage of the NAM scheme is that it guarantees convergence for any choice of the Lagrange multipliers $(s_D, s_P)$, without requiring additional assumptions. However, the use of Newton’s method entails differentiability of the perceptual metric and introduces additional computational complexity at each iteration. The RAM scheme, by contrast, eliminates the differentiability requirement and avoids the extra per-iteration computational cost, but at the price of restricting the range of $(s_D, s_P)$ values for which convergence to the optimal solution is guaranteed. As a consequence, this limitation may prevent the computation of the full RDPF curve.
\end{remark}
\par
\begin{figure}[!t]
    \centering
    
     \subfloat[NAM scheme ]{%
        \includegraphics[width=0.47\linewidth]{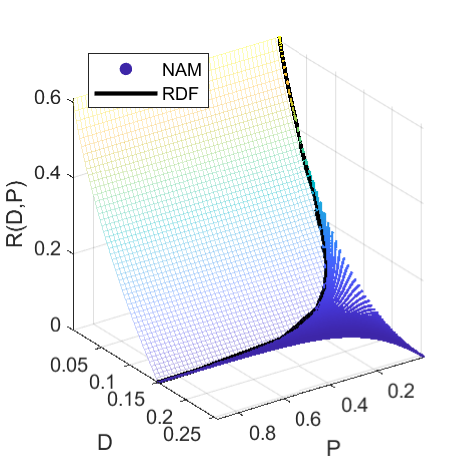}
        \label{fig:KL_NAM}
    }
    \hfill
    \subfloat[RAM scheme ]{%
        \includegraphics[width=0.47\linewidth]{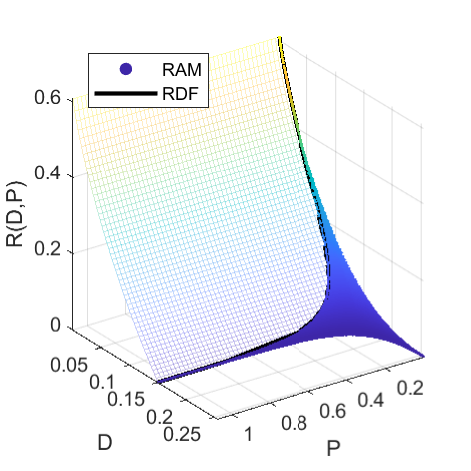}
        \label{fig:KL_RAM}
    }
   \caption{$R(D,P)$ for a Bernoulli source under Hamming distortion and $D_f(\cdot||\cdot)=D_{KL}(\cdot||\cdot)$.}\label{fig:1}
    \end{figure}

 In Fig. \ref{fig:1}, we illustrate the RDPF trade-offs of \eqref{eq.sec2:rdpf} using algorithmic implementations based on NAM and RAM methods. The results are obtained for ${\mathcal X} = \hat{\mathcal{X}}=\{0,1\}$ with $\p{} \sim Ber(0.15)$, with $\Delta(\cdot,\cdot) = \Delta_H(\cdot,\cdot)$ and perception constraint chosen to be $D_f(\cdot||\cdot) = D_{KL}(\cdot||\cdot)$.

\paragraph*{The notable case of TV perception constraint} TV distance is one of the most interesting $f$-divergences because it is not differentiable everywhere, particularly at points where its graph has sharp changes in slope, such as corners, cusps, or discontinuities. The non-differentiability is primarily due to the use of an absolute value ($L_1$ norm) in its definition, which results in a ``V-shape" in the derivative calculation at zero. 
\par
\begin{figure}[ht]
\vspace{-0.8cm}
    \centering
    \subfloat[$D_f(\cdot||\cdot)=TV(\cdot||\cdot)$]{%
        \includegraphics[width=0.47\linewidth]{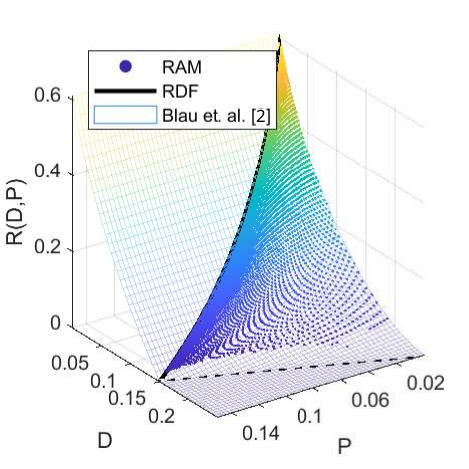}
        \label{fig:2a}
    }
    \hfill
    \subfloat[$n = 1$]{%
        \includegraphics[width=0.47\linewidth]{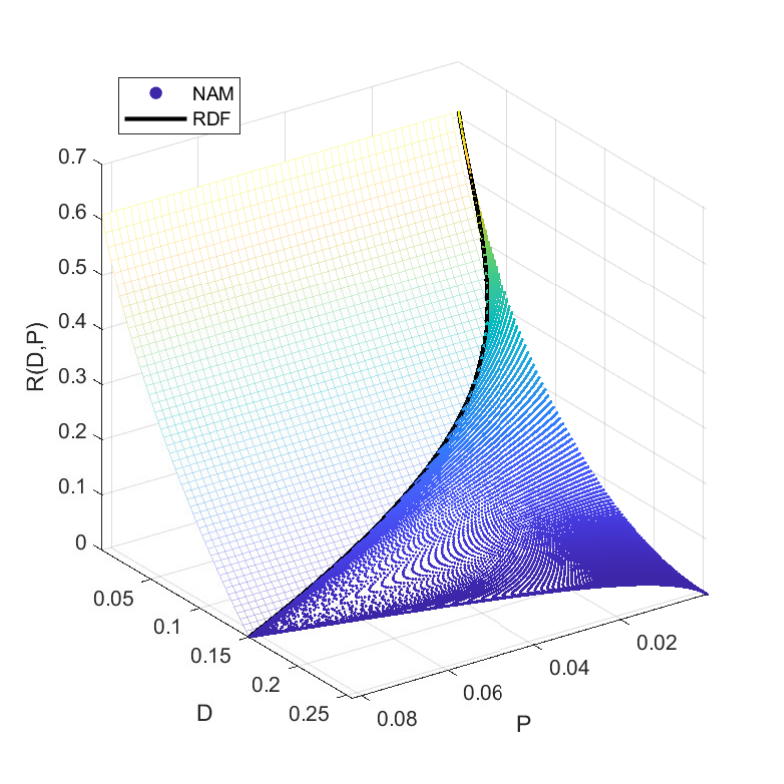}
        \label{fig:2b}
    }

    \subfloat[$n = 10$]{%
        \includegraphics[width=0.47\linewidth]{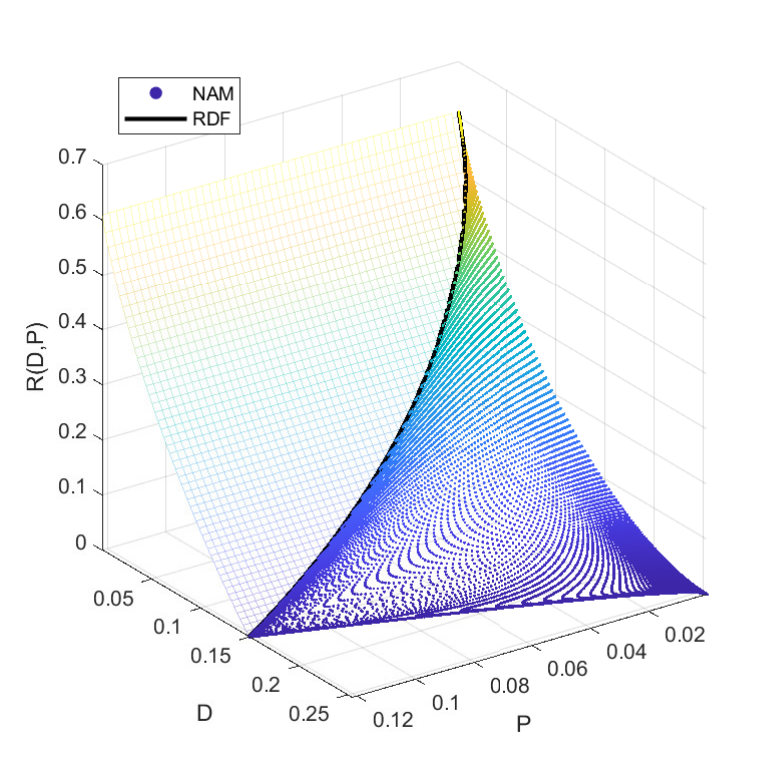}
        \label{fig:2c}
    }
    \hfill
    \subfloat[$n = 100$]{%
        \includegraphics[width=0.47\linewidth]{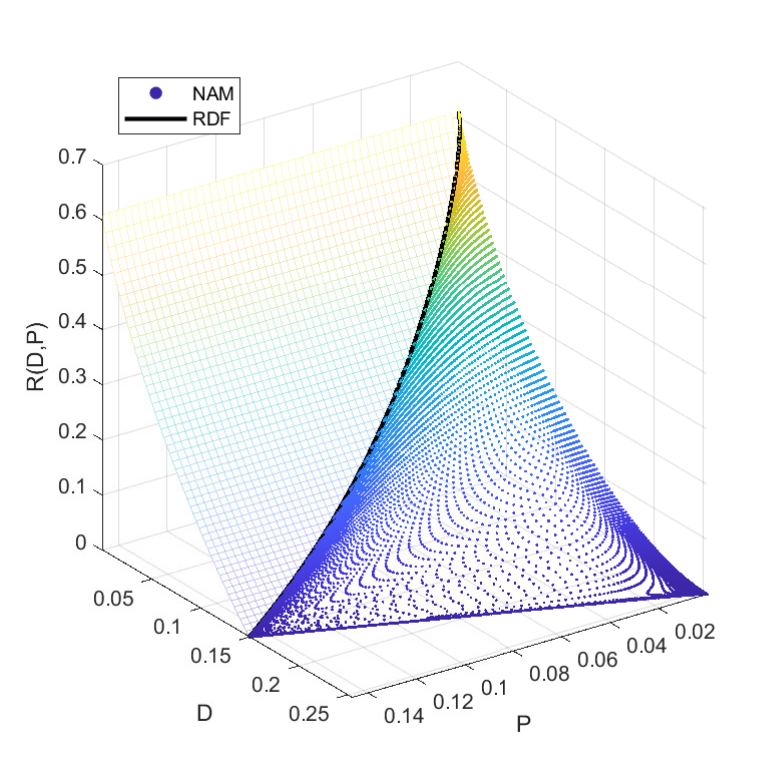}
        \label{fig:2d}
    }

    \caption{RDPF under Hamming distortion and $TV$ perception, computed with the RAM scheme, and under an approximation of TV via the sequence $\{D_{f_n}(\cdot||\cdot):~n=1,2,\ldots\}$, computed with the NAM scheme, for $n \in \{1, 10, 100\}$.}
    \label{fig:2}
    \vspace{-0.3cm}
\end{figure}

 Due to the functional properties of the TV distance, the NAM method is not directly applicable using \cite[Algorithm 1]{serra:2025tit}. Moreover, the RAM method implemented via \cite[Algorithm 2]{serra:2025tit} does not fully recover the RDPF trade-offs, owing to its inherent limitation of shrinking the feasible range of $(s_D, s_P)$. This limitation becomes evident when comparing the numerical computation for a Bernoulli source with Hamming distance against the optimal analytical expression derived in \cite[Equation 6]{blau:2019}, as shown in Fig. \ref{fig:2a}.
\par To address this issue, we approximate the $TV(\cdot\|\cdot)$ distance by a sequence of $f$-divergences $\{D_{f_n}(\cdot\|\cdot) : n = 1,2,\ldots\}$ such that $D_{f_n}(\cdot\|\cdot) \to TV(\cdot\|\cdot)$ as $n \to \infty$. Specifically, we consider a family of convex functions of the form
\begin{equation}
f_n(x) = \frac{2}{\pi}(x-1)\arctan\bigl(n(x-1)\bigr),\nonumber
\end{equation}
and characterize the corresponding sequence of $f$-divergences $\{D_{f_n}(\cdot\|\cdot):~n=1,2,\ldots\}$, which converges uniformly to $TV(\cdot\|\cdot)$ as $n \to \infty$ and satisfies $D_{f_n}(\cdot\|\cdot) \le TV(\cdot\|\cdot)$ (see \cite[Lemmas 7, 8]{serra:2025tit}).
\par The latter implies that the RDP problem defined with the perception metric $D_{f_n}(\cdot||\cdot)$ constitutes a lower bound on the RDP problem defined with the $TV(\cdot||\cdot)$ perception metric. 
Furthermore, since $D_{f_n}(\cdot||\cdot)$ is a smooth function, the NAM method can be applied to compute the corresponding RDPF.
Fig. \ref{fig:2b}-\ref{fig:2d} depict the RDPF for a Bernoulli source under Hamming distortion and the perception measure $D_{f_n}$, for $n \in \{ 1, 10, 100 \}$. The results show that, as $n$ increases, $D_{f_n}$ provides an increasingly accurate approximation of the $TV(\cdot||\cdot)$ distance.

\subsection{Continuous Sources under Perceptual Constraints and Perfect Realism}\label{section:iiib}

In this section, we review the computation of \eqref{eq.sec2:rdpf} for continuous sources, with particular emphasis on Gaussian sources. We focus on the case where the distortion measure is the mean-squared error (MSE), and the perception constraint is drawn from the family of $\alpha$-divergences or the squared Wasserstein-2 distance. We also consider the special case of perfect realism.

\paragraph*{Gaussian sources under $\alpha$-divergence and squared Wasserstein-2 distance} Here, we present a generic method for computing \eqref{eq.sec2:rdpf} when $\Delta(x,\hat{x})=|x-\hat{x}|^2$ and $D(p_X,p_{\hat{X}})=D_\alpha(\cdot||\cdot)$. The $\alpha$-divergences, first proposed by Chernoff \cite{Chernoff:52} and later studied extensively by Amari \cite{Amari:85}, form a canonical family of divergences at the intersection of the class of $f$-divergences and that of Bregman divergences on the manifold of positive measures. Since $\alpha$-divergences are convex in both arguments, the resulting RDPF is a convex optimization problem. For general alphabets, $\alpha$-divergences admit several equivalent representations, such as Csiszár’s form in \eqref{cziscar_a_divergence}. In the sequel, however, we adopt a variant of Amari’s formulation, namely,
\begin{align}
    D_\alpha(p_X||q_{\hat{X}}) = \frac{1}{\alpha(\alpha - 1)} \left( \int_{-\infty}^{\infty} p_X^{\alpha} q_{\hat{X}}^{1-\alpha} \,dx - 1 \right),
    \label{eq:Amari}
\end{align}
for $\alpha \in \mathbb{R} \backslash \{0, 1\}$.
\par
\begin{figure}[htbp]
\centerline{\includegraphics[width=0.35\textwidth]{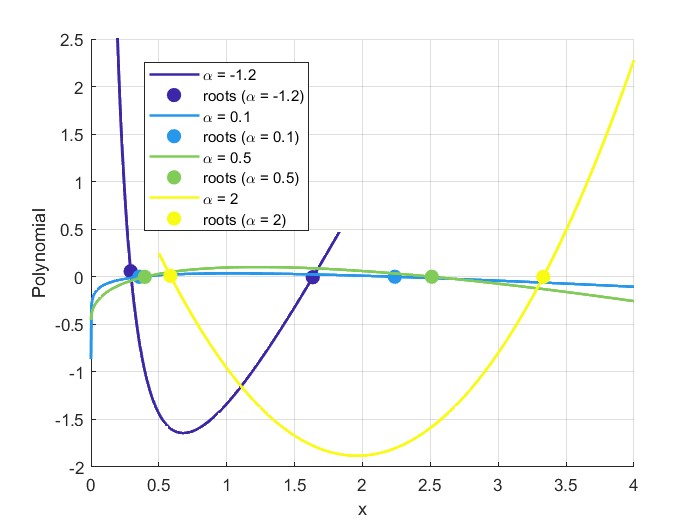}}
    \caption{Polynomial \eqref{two_root_equation} for  $\alpha = -1.2$, $\alpha = 0.1$, $\alpha = 0.5$, and $\alpha = 2$, with the perception constraint fixed to $P = 0.2$.}
    \label{fig:3}
\end{figure}

 Without loss of optimality \cite[Lemma 2]{sourla:2024}, we assume scalar-valued $X\sim{p}_X={\cal N}(0,\sigma_{X}^2)$ and scalar-valued $\hat{X}\sim{p}_{\hat{X}}={\cal N}(0,\rho_{\hat{X}}^2)$. Due to the monotonicity of the resulting RDPF optimization problem, it is natural to focus on the case where the perception constraint $D_{\alpha}(p_X\|q_{\hat{X}}) = P$. The latter can be expressed as a polynomial equation of the form 
\begin{align}
g(x) = x^{\alpha} - \alpha C x - (1 - \alpha)C=0\label{two_root_equation}    
\end{align}
where $x = \frac{\rho^2}{\sigma^2}$ and $C = (1 - \alpha(1-\alpha)P)^2$. Furthermore, for all $\alpha \in (-\infty, 0) \cup (0,1) \cup (1, +\infty)$, \eqref{two_root_equation} has two roots, $r_0$ and $r_1$, such that $r_0 \in [0, x_0]$ and $r_1 \in [x_0, y_0]$, where $x_0$ is the unique stationary point of \eqref{two_root_equation} and $y_0 = (x_0 + \epsilon) - \frac{f(x_0 + \epsilon)}{f'(x_0 + \epsilon)}$ for $\epsilon > 0$. An illustrative example of a particular numerical simulation where \eqref{two_root_equation} is solved using the bisection method is shown in Fig. \ref{fig:3}.
The following result, which can be found in \cite[Theorem 1]{sourla:2024}, derives a parametric upper-bound expression for \eqref{eq.sec2:rdpf}, hereinafter denoted by $R^G(D, P)$, under the assumption of jointly Gaussian sources. This upper bound coincides with the exact value of $R(D, P)$ when the $\alpha$-divergence reduces to the reverse KL distance.
\begin{theorem}\cite{sourla:2024} \textit{(Parametric solution of $R^G(D, P)$)}\label{thm:gaussian_rdp:alpha_div}
Let $X$ be a scalar Gaussian source $X \sim \mathcal{N}(0,\sigma_{X}^2)$. Then, $R^G(D,P)$ under an MSE distortion and $\alpha$-divergence perception is achieved by a jointly Gaussian reconstruction $\hat{X} \sim \mathcal{N}(0,\rho_{\hat{X}}^2)$ and is given by
\begin{align}
    R^G(D, P)
    &= \begin{cases}
        \max\left\{\frac{1}{2}\log\frac{\sigma_{X}^2}{D}, 0\right\} 
            &~\textit{if } (D,P) \in \mathcal{S} \\ 
        \frac{1}{2}\log\frac{2 \rho_{\hat{X}}^2\sigma_X^2}{\rho^2\sigma^2 - \left( \frac{\sigma_X^2 + \rho_{\hat{X}}^2 - D}{2} \right)^2}  &~\textit{if } (D,P) \notin \mathcal{S} 
    \end{cases}\label{eq: RDPF_alpha_closed}
\end{align}
where
\begin{align*}
    \rho^2
    &= \begin{cases}
        \sigma^2 - D &\quad\textit{if } (D,P) \in \mathcal{S} \\ 
        \min\{r_0,r_1\}  &\quad\textit{if } (D,P) \notin \mathcal{S}. 
    \end{cases} \nonumber\\
    \mathcal{S} &= \Big\{(D,P) \in \mathbb{R}_+^2: P > g(D,\sigma_X)~~\land~~\\
    & \qquad (\alpha - 1)\left(\left|1 - \frac{D}{\sigma_X^2}\right| - \left(1 - \frac{1}{\alpha}\right) \right) > 0 \Big\}\\
    &g(D,\sigma_X) \nonumber\\
    &= \frac{1}{\alpha(1-\alpha)}\left(1-\frac{\sigma_X^{1-\alpha}|\sigma_X^2-D|^{\alpha/2}}{\sqrt{\alpha|\sigma_X^2-D|+(1-\alpha)\sigma_X^2}}\right)
\end{align*}
with $r_0$ and $r_1$ being the roots of \eqref{two_root_equation}.
\end{theorem}
Theorem~\ref{thm:gaussian_rdp:alpha_div} is quite general and recovers analytical expressions for several known $\alpha$-divergences \cite{serra:2024}. While an extension of Theorem~\ref{thm:gaussian_rdp:alpha_div} to multivariate Gaussian sources is, in principle, possible, it is computationally intensive, as it requires solving the polynomial equation in~\eqref{two_root_equation} in matrix form.
\par An alternative way to tackle the problem at hand is to adopt a different perspective. Instead of following this direction, in \cite{serra:2024} we propose a \emph{divergence-specific approach}, in which a particular $\alpha$-divergence, or another suitable divergence that admits an analytical expression under jointly Gaussian random variables, is selected and the resulting problem is addressed directly.
To make this point clear, we focus on the squared Wasserstein-2 distance as a representative example because it preserves Gaussianity on the reconstruction side $\hX$ similar to the reverse KL divergence, and thus achieves $R(D, P)=R^G(D, P)$. Recall that the squared Wasserstein-2 distance is defined as
\begin{align}
\WTS(\pdf{X},\pdf{\hX}) \triangleq \min_{\Pi} \E{|X - \hX|^2}\label{exp:wasserstein-2}
\end{align}
where $\Pi$ denotes the set of all joint distributions $p_{X,\hX}$ with the prescribed marginals $\pdf{X}$ and $\pdf{\hX}$.
\begin{theorem} {\cite{serra:2024}} \label{thm:uRDP:W2}
Consider the optimization problem in \eqref{eq.sec2:rdpf} for jointly Gaussian $(X,\hX)$ with $D(\cdot||\cdot)\equiv{W}_2^2(\cdot,\cdot)$. Then $R^G(D,P)$ can be obtained analytically as follows:
\begin{align}
    \begin{split}
        &\RG(D,P) \\
        & = \begin{cases}
                \max \left\{  \frac{1}{2} \log \left( \frac{\sigma_X^2}{D} \right), 0 \right\}~~\text{if } (D,P) \in \mathcal{S}^c\\
                 \frac{1}{2} \log \left( 1 +  \frac{\left[ \sigma_X^2 + (\sigma_X - \sqrt{P})^2 - D\right]^2}{(D - P) \left[(2\sigma_X - \sqrt{P})^2 - D\right]} \right)~~\text{if } (D,P) \in \mathcal{S}\\
        \end{cases} \label{eq:RDP_W2}
    \end{split}
\end{align}
where  $\mathcal{S} = \left\{ (D,P): \sqrt{P} \le \sigma_X - \sqrt{|\sigma_X^2 - D|} \right\}$. Moreover, the solution that achieves $R^G(D, P)$ is realized by a linear model $\hat{X} = aX + W$, where the design variables $(a,\sigma_W^2)$ are obtained in closed form as follows:
\begin{align}
    \begin{split}
        a= 
        \begin{cases}
            \max \left\{1 - \frac{D}{\sigma_W^2}, 0 \right\} \quad \text{if } (D,P) \in \mathcal{S}^c\\
             \frac{1}{2} \frac{\sigma_X^2 + \left(\sigma_X - \sqrt{P} \right)^2 - D}{\sigma_X^2} \quad \text{if } (D,P) \in \mathcal{S}\\
        \end{cases}\\
        \sigma_W^2 = \min\{D, \sigma_X^2 + (\sigma_X - \sqrt{P})^2\} - \left( 1 - {a}\right)^2\sigma_X^2.
    \end{split} \label{design_variables_wesserstein}
\end{align}
\end{theorem}
{We emphasize the following remark on Theorem \ref{thm:uRDP:W2}.
\begin{remark}(On Theorem \ref{thm:uRDP:W2})
The closed-form expression of $R^G(D,P)$ in \eqref{eq:RDP_W2} for the squared Wasserstein-2 distance, $W^2_2(\cdot,\cdot)$. was first derived in \cite[Theorem~1]{zhang:2021}. However, \cite{zhang:2021} does not explicitly derive the forward realization/test channel that attains this value. Here, we complement that characterization by deriving the linear realization $\hat{X} = a{X} + W$ together with the closed-form design parameters $a$ and $\sigma_{W}^2$ in \eqref{design_variables_wesserstein}. This constructive characterization is important because it identifies the corresponding optimal linear encoder-decoder structure and connects the information-theoretic limit to implementable coding schemes such as lattice coding \cite{zamir:2014}, with a relatively small gap between the operational and information-theoretic limits.    
\end{remark}
}
To derive the result of Theorem \ref{thm:uRDP:W2}, we rely on two key ingredients. First, under the assumption of joint Gaussianity, the optimal minimizer (or test channel) admits a forward realization \cite{berger:1971} that achieves the Gaussian RDPF in close analogy to the classical RD setting. Second, to determine the design parameters of this forward realization, namely, 
$(a,\sigma_W^2)$, we invoke the Karush--Kuhn--Tucker (KKT) conditions, leveraging the convexity of the underlying optimization problem. \\
\noindent{\bf Multivariate Gaussian Random Vectors.} In what follows, we show how to derive, under certain conditions, a generic alternating minimization approach for computing \eqref{eq.sec2:rdpf} for multidimensional Gaussian random sources with jointly Gaussian reconstruction. To this end, we first introduce a convenient parametric representation of $R^G(D,P)$ as follows.
\par Given a Gaussian source $X \sim {\cal N}(0, \Sigma_X)$, with $\Sigma_X \succ 0$, assume that the reconstructed random vector $\hat{X}\in\mathbb{R}^N$ is chosen such that the joint tuple $(X,\hat{X})$ is jointly Gaussian. Then, the reconstructed message admits a linear (forward) realization of the form $\hat{X} = AX + W$, where $A\in\mathbb{R}^{N\times{N}}$, $W \sim {\cal N}(0, \Sigma_W)$, $W \perp X$ and $\Sigma_W \succeq 0$. Moreover, we can cast \eqref{eq.sec2:rdpf}  as follows:
\begin{align}
    \begin{split}
        R(D,P) &\le \RG(D,P) \\
        &= \min_{\substack{A\in\mathbb{R}^{N\times{N}},\Sigma_W\succeq{0}\\\E{\Delta(X,\hX)} \le D\\ D(\pdf{X}||\pdf{\hX}) \le P
        }}  \frac{1}{2} \log \left( \frac{|A\Sigma_X{A}^T + \Sigma_W|}{|\Sigma_W|} \right)
    \end{split} \label{opt: generic_gaussian_vector}
\end{align}
where the specific form of the fidelity constraints $\E{\Delta(\cdot,\cdot)}$ and $D(\cdot||\cdot)$ is not yet specified.

Under the assumption that the fidelity constraints $\E{\Delta(\cdot,\cdot)}$ and $D(\cdot||\cdot)$ are tensorizable, i.e., $\E{\Delta(X,\hX)} \ge \sum_{i = 1}^N g\left(\E{\Delta(X_i,\hX_i)}\right)$ and $ D(\pdf{X}||\pdf{\hX}) \ge \sum_{i = 1}^N h\left(D(\pdf{X_i}||\pdf{\hX_i})\right)$, with $g(\cdot)$ and $h(\cdot)$ being convex functions, then we can apply \cite[Lemma 2]{serra:2024} to \eqref{opt: generic_gaussian_vector} to obtain the following lower bound:
\begin{align}
        &\RG(D,P) \nonumber\\
        &\stackrel{(\star)}\ge  \min_{\substack{\eigvi{A}, \eigvi{\Sigma_W}\\ \sum_{i=1}^N g\left(\E{\Delta(X_i,\hat{X}_i)}\right) \le D \\ h\left(D(\pdf{X_i}||\pdf{\hat{X}_i})\right) \le P} } \sum_{i=1}^N  \frac{1}{2}\log\left(1+ \frac{\lambda_{A,i}^2\eigvi{\Sigma_X}}{\eigvi{\Sigma_W}}\right)
   \label{opt: RDPFonTheEigenvectors}
\end{align}
where $(\star)$ holds with equality if the triplet $(A,\Sigma_W, \Sigma_X)$ commutes pairwise. In fact, for jointly Gaussian random vectors $(X,\hX)$, the sufficient condition for achieving the lower bound in \eqref{opt: RDPFonTheEigenvectors} can easily be realized, since the matrices $(A, \Sigma_W)$ are design variables and can be chosen to share the \textit{same eigenvectors} as $\Sigma_X$ \cite[Proposition 1]{stavrou:2022tac}. As a result, one can replace the inequality by equality without loss of generality. We note that the inequality in  \eqref{opt: RDPFonTheEigenvectors} does not hold with equality, in general, beyond $\IID$ random vectors.\\
\noindent{\bf Proposed minimization method.} To solve \eqref{opt: RDPFonTheEigenvectors}, we first introduce the (vector) optimization variables $\vecD = [D_i]_{i \in \{1,\ldots,N\}}$ and $\vecP = [P_i]_{i \in \{1,\ldots,N\}}$ as 
\begin{align*}
D_i = \E{\Delta(X,\hat{X}_i)},~P_i = D(\pdf{X_i}||\pdf{\hat{X}_i})~i\in\{1,\ldots,N\}.
\end{align*}
Once the slack variables above are substituted into \eqref{opt: RDPFonTheEigenvectors}, we obtain
\begin{align}
        \RG(D,P) &= \min_{\substack{\vecD, \vecP\\\sum_{i=1}^N g(D_i) \le D\\ \sum_{i=1}^N h(P_i) \le P }} \sum_{i=1}^N \RG_i(D_i,P_i)  \label{opt:gmrdpf:full}
\end{align}
where $\RG_i(D_i,P_i)$ corresponds to the stagewise problem
\begin{align}
       \RG_i(D_i,P_i) &= \min_{\substack{\eigvi{A}, \eigvi{\Sigma_W}\\ D_i=\E{\Delta(X_i,\hat{X}_i)}\\ P_i=D(\pdf{X_i}||\pdf{\hat{X}_i})}}  \frac{1}{2}\log\left(1+ \frac{\lambda_{A,i}^2\eigvi{\Sigma_X}}{\eigvi{\Sigma_W}}\right).\nonumber
\end{align}
Note that \eqref{opt:gmrdpf:full} gives rise to three distinct rate-region cases, which together characterize the full Gaussian RDPF. In particular, we may have: (i) only the distortion constraint is active (hereinafter referred to as \textbf{Case I}), (ii) only the perception constraint is active (hereinafter referred to as \textbf{Case II}), or (iii) both constraints are active (hereinafter referred to as \textbf{Case III}). Notably, \textbf{Case III} is the most relevant scenario since the computation of the other two cases can be recovered as special instances of it.
\par {To find the optimal pair $(\vecD^{*},\vecP^{*})$ in \eqref{opt:gmrdpf:full} we employ an alternating minimization technique. Specifically, we decompose the problem \eqref{opt:gmrdpf:full} into the following two subproblems
\begin{itemize}
    \item For fixed $\vecP$, \eqref{opt:gmrdpf:full} simplifies to
        \begin{align}
               \min_{\substack{\vecD\\\sum_{i=1}^N g(D_i) \le D}} \sum_{i=1}^N \RGi(D_i,P_i). \label{opt:gmrdpf:partial_optD}
        \end{align} 
    \item For fixed $\vecD$, \eqref{opt:gmrdpf:full} simplifies to
    \begin{align}
           \min_{\substack{\vecP\\\sum_{i=1}^N h(P_i) \le P}} \sum_{i=1}^N \RGi(D_i,P_i) .  \label{opt:gmrdpf:partial_optP}
    \end{align}
\end{itemize}
}
The methodology described above is commonly referred to in the literature as the \textit{block nonlinear Gauss–Seidel method} \cite{grippo:2000}. We next state a convergence result for the resulting iterative procedure.
\begin{theorem} \cite{serra:2024} (Convergence) \label{th:gmrdpf:RDPF_AM}
    Let the optimization problem \eqref{opt:gmrdpf:full} be defined for finite distortion and perception levels $(D,P)$. Let  $(\vecD^{(0)},\vecP^{(0)})$ be an initial point, and let the sequence $\{(\vecD^{(n)},\vecP^{(n)}):~n=1,2,\ldots\}$ be obtained by alternately optimizing \eqref{opt:gmrdpf:partial_optD} and \eqref{opt:gmrdpf:partial_optP}. Then the sequence has a limit $\lim_{n \to \infty}(\vecD^{(n)},\vecP^{(n)})  = (\vecD^{*},\vecP^{*})$, and this limit is an optimal solution of \eqref{opt:gmrdpf:full}. 
\end{theorem}
To solve the constrained optimization problems in \eqref{opt:gmrdpf:partial_optD} and \eqref{opt:gmrdpf:partial_optP}, we can introduce Lagrange multipliers to form equivalent unconstrained (parametric) problems. Substituting these into \eqref{opt:gmrdpf:full} yields a fully parametric, unconstrained formulation. 
Let $s = (s_D, s_P)$, with $s_D > 0$ and $s_P > 0$, be the vector of Lagrange multipliers associated with the distortion (i.e., $s_D$) and perception (i.e., $s_P$) constraints, respectively. Then, the Lagrangian functional $L_{\RG}(s)$ associated with \eqref{opt:gmrdpf:full} is defined as follows:
\begin{align}
    \begin{split}
     \min_{\vecD, \vecP} &L_{\RG}(\vecD, \vecP, s) \triangleq \min_{\vecD, \vecP}\Bigg[ \sum_{i=1}^N \RG_i(D_i,P_i)  \\
     &+ s_D \sum_{i=1}^N g(D_i) + s_P\sum_{i=1}^N h(P_i)\Bigg]. \end{split}\label{opt:gmrdpf:fullLagrangian}
\end{align}

Similarly to the constrained case, the optimal pair $(\vecD^{*},\vecP^{*})$ in \eqref{opt:gmrdpf:fullLagrangian} can be characterized through an alternating minimization scheme. Hence, the associated subproblems are as follows:
\begin{itemize}
    \item For fixed $\vecP$,
        \begin{align}
            \begin{split}
               \min_{\vecD} \sum_{i=1}^N \RG_i(D_i,P_i) + s_D \sum_{i=1}^N g(D_i).
           \end{split} \label{opt:gmrdpf:partial_optD_Lagrangian}
        \end{align} 
    \item For fixed $\vecD$,
    \begin{align}
        \begin{split}
           \min_{\vecP} \sum_{i=1}^N \RG_i(D_i,P_i) + s_P\sum_{i=1}^N h(P_i).
       \end{split} \label{opt:gmrdpf:partial_optP_Lagrangian}
    \end{align}
\end{itemize}
Assume the Lagrange multiplier vector $s$ is given and let $\vecD_s^*$ and $\vecP_s^*$ be the optimal solutions obtained from the Gauss-Seidel method by alternating solving \eqref{opt:gmrdpf:partial_optD_Lagrangian} and \eqref{opt:gmrdpf:partial_optP_Lagrangian}, respectively. Furthermore, let $D_s = \sum_{i=1}^N g(D^*_{s,i})$ and $P_s = \sum_{i=1}^N h(P^*_{s,i})$. Then, leveraging Lagrangian duality, we can compute $\RG(D_s, P_s)$ as
\begin{align}
    \RG(D_s, P_s) = L_{\RG}(\vecD_s^*,\vecP_s^*,s) - s_D D_s - s_P P_s.
\end{align}

This reformulated problem can then be solved efficiently using the algorithm in \cite[Algorithm 1]{serra:2024}, which is guaranteed to converge at a sublinear rate (at worst).
\noindent\textbf{{Generality of the approach and comparison to related literature.}} The approach outlined above, together with its algorithmic implementation in \cite{serra:2024}, can optimally solve any problem of the form \eqref{opt: generic_gaussian_vector}, provided that the perception constraint admits an analytical characterization for jointly Gaussian random variables. {As a result, this approach is more general than, for example, the approach adopted in \cite{qian:2025}, which aims to derive analytical generalized reverse water-filling solutions for \textit{divergence-specific} problems}.
\begin{figure}[t]
        \centering
\includegraphics[width = 0.9\linewidth]{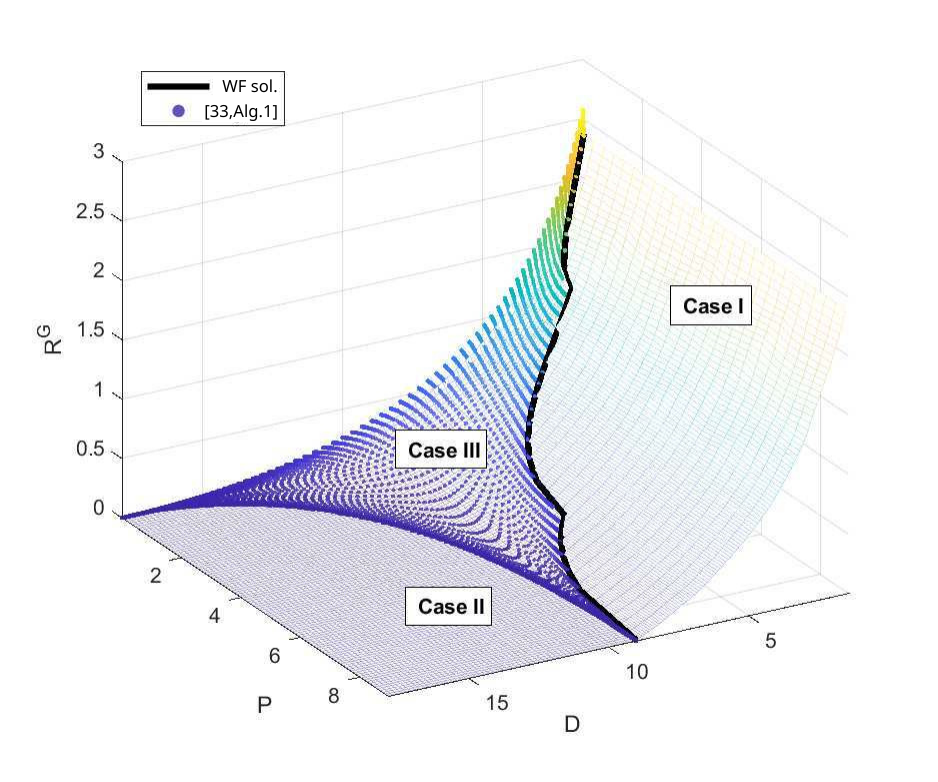}
\caption{$\RG(D,P)$ for a Gaussian source $X \sim \ND(0,\Sigma_X)$ with $\Sigma_X = \diag([1,3,5])$ under a squared Wasserstein-2 distance.} \label{fig:4}
\end{figure}
\par In Fig.~\ref{fig:4}, we use \cite[Algorithm 1]{serra:2024} to compare $\RG(D,P)$ under the squared Wasserstein-2 perception measure with the solution of the classical RDF problem obtained via the reverse water-filling algorithm (black line). For the classical case, the perception metric is computed \emph{a posteriori} using the same divergence measure. The results confirm that, for a bounded divergence metric, the classical rate--distortion solution arises as an extreme case of the $\RG(D,P)$ surface, thereby recovering the boundary separating \textbf{Case I} and \textbf{Case III}. Moreover, the region corresponding to \textbf{Case I} can be obtained by a rigid translation of this boundary curve, since any $(D,P)$ point in this region defines a Gaussian RDPF problem in which the perception constraint is inactive, and therefore equivalent to the classical RDF problem.  
\par
\begin{figure}[htp]
    \centering
    \includegraphics[width = \linewidth]
    {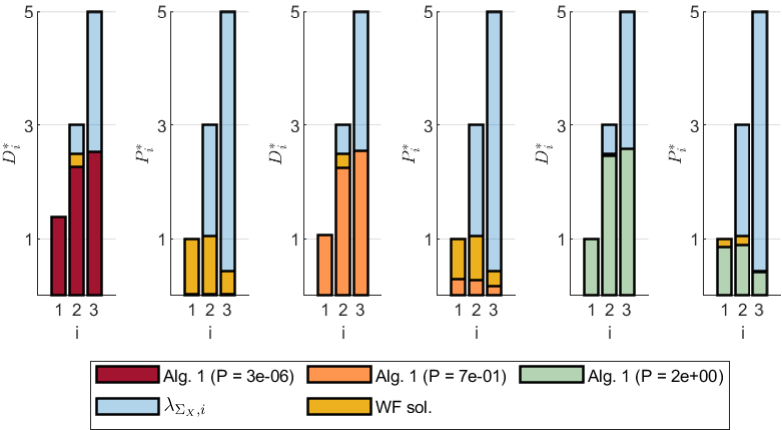}
    \caption{ Comparison of the per-dimension distortion $D_i^*$ and perception levels $P_i^*$ for a fixed target distortion level $D = 6$ between the reverse water-filling solution and \cite[Algorithm 1]{serra:2024}. }
    \label{fig:5}
\end{figure}

 In Fig. \ref{fig:5}, we illustrate how the optimal per-dimension distortion and perception levels, $D_i^*$ and perception $P_i^*$, distribute across the source components when reconstructing $X$ into $\hat{X}$. In particular, we compare the solution obtained via \cite[Algorithm 1]{serra:2024} against the classical RD reverse water-filling allocation. Throughout the simulation study, we fix the target distortion to $D = 6$ and vary the target perception constraint $P$. The most stringent regime ($P \approx 0$) provides a representative example of the \emph{perfect realism} behavior: rather than following a single uniform water level, the optimal distortions $D_i^*$ adapt to each marginal component (as discussed in more detail in the sequel). A similar, though less pronounced, trend is observed for the intermediate settings ($P \approx 0.7$ and $P \approx 2$), where relaxing the perception constraint leads to a more uniform allocation of $D_i^*$ across dimensions. This gradual transition highlights how looser perception requirements induce behavior increasingly aligned with the classical reverse water-filling solution.

\paragraph*{Multivariate Gaussian sources under the perfect realism regime} In what follows, we specialize the previous results for the multivariate jointly Gaussian setting to the perfect realism scenario, i.e., the case in which \eqref{opt: generic_gaussian_vector} satisfies $p_X = p_{\hat{X}}$ almost surely (a.s.), and therefore $P=0$. We omit the scalar jointly Gaussian case, since it follows directly as a special case of the multivariate setting.
\par In the following theorem, we state the characterization of the optimal solution to the subproblem \eqref{opt:gmrdpf:partial_optD_Lagrangian}.

\begin{theorem} \cite{serra:2024} \label{th:gmrdpf:AM_subD}
Let the Lagrange multiplier $s_D > 0$ be given. Then, for fixed $\vecP$, the optimal stagewise distortion levels $\vecD^*(\vecP) = [D^*_i(P_i)]_{i \in \{1,\ldots,N\}} \in \mathcal{S}$ achieving the minimum of \eqref{opt:gmrdpf:partial_optD_Lagrangian} are
    \begin{align}
        \begin{split}
            D^*_i &=  P_i + 2 \sqrt{\lambda_{\Sigma_X,i}} \left(\sqrt{\lambda_{\Sigma_X,i}} - \sqrt{P_i} \right) + \frac{1}{2s_D} \\
            & \quad - \sqrt{4\eigvi{\Sigma_X}(\sqrt{\lambda_{\Sigma_X,i}} - \sqrt{P_i})^2 + \frac{1}{4s_D^2}} . 
        \end{split} \label{eq:gmrdpf:AM_subD:optD}
    \end{align}
\end{theorem}
{
The expression in Theorem \ref{th:gmrdpf:AM_subD} also admits a useful reverse water-filling interpretation, which clarifies how the perception constraint modifies the classical reverse water-filling solution for Gaussian vector sources \cite{berger:1971}. This is explained in the next remark.}
{
\begin{remark}\label{remark:reverse_water_vector}(Adaptive water levels for fixed perception allocation) The allocation in Theorem \ref{th:gmrdpf:AM_subD} can be interpreted as a perception-constrained analogue of reverse water-filling. In the classical multivariate Gaussian RDF, the Lagrange multiplier $s_D$ induces a single dimension-independent water level $w(s_D)=\frac{1}{2s_D}$, and the optimal distortion is clipped at the source eigenvalue $\lambda_{\Sigma_X,i}$. By contrast, in Theorem \ref{th:gmrdpf:AM_subD} the effective allocation depends jointly on the source eigenvalue $\lambda_{\Sigma_X,i}$ and on the prescribed per-component perception level $P_i$. Thus, the common multiplier $s_D$ does not generate a uniform water level across all dimensions; instead, it induces component-dependent adaptive water levels whose shape is determined by both the variance profile of the source and the perception allocation. This behavior is consistent with the generalized reverse water-filling interpretation developed for Gaussian vector RDP problems under KL-divergence and squared Wasserstein-2 perception constraints in \cite{qian:2025}, where active distortion and perception constraints lead to positive rate allocation across all components and generally unequal effective water levels. As the perception constraint is relaxed, the allocation progressively approaches the classical reverse water-filling behavior. In contrast, stringent perception constraints force the reconstruction to preserve the source distribution more faithfully across all components, as illustrated in Fig. \ref{fig:5}.
\end{remark}
}
The results of Theorem \ref{th:gmrdpf:AM_subD} characterize the optimal distortion vector $\vecD$ for a given perception vector $\vecP$. As an additional result, Theorem \ref{th:gmrdpf:AM_subD} gives the closed-form solution for the case where $\vecP = \mathbf{0}$ (the all-zeros vector), or equivalently $d(\pdf{X}||\pdf{\hX}) = 0$, referred to as {\it perfect realism} \cite{blau:2019,chen:2022}.
\begin{corollary} \cite{serra:2024} \label{th: perfect_realism}
    Consider the optimization problem \eqref{opt:gmrdpf:fullLagrangian} for perception level $\vecP = \mathbf{0}$. Then, for a given Lagrange multiplier $s_D > 0$, the optimal solution $\vecD^* = [D_i^*]_{i \in 1,\ldots,N}$ is given by
    \begin{align}
        D^*_i = 2 \eigvi{\Sigma_X} + \frac{1}{2s_D} - \sqrt{4\eigviSq{\Sigma_X} + \frac{1}{4s_D^2}} \label{eq:perfect_realism_solution}
    \end{align} 
   such that the distortion level $D = \sum_{i = 1}^N D^*_i$.
\end{corollary}
{
\begin{remark}\label{remark:perfect_real_water}(On Corollary \ref{th: perfect_realism}) Corollary \ref{th: perfect_realism} is the perfect-realism specialization of the adaptive allocation in Theorem \ref{th:gmrdpf:AM_subD}, obtained by setting $P_i=0$ for all components. In this case, the optimal distortion allocation is given by \eqref{eq:perfect_realism_solution}. Hence, $D^*_i\rightarrow{0}$ as $s_D\rightarrow\infty$, while $D_i^*\rightarrow2\lambda_{\Sigma_X,i}$ as $s_D\rightarrow{0}$. This differs sharply from the classical multivariate Gaussian RDF, where the reverse water-filling solution is 
\begin{align*}
    D^*_{i,RD} = \min\left( w(s_D), \eigvi{\Sigma_X} \right) \qquad w(s_D) = \frac{1}{2s_D}
\end{align*}
and the water level is dimension independent. Under perfect realism, a component cannot simply be removed from the reconstruction distribution: even when its mutual-information contribution becomes small, the reconstruction must still preserve the correct marginal law. This explains why the effective distortion ceiling is $2\eigvi{\Sigma_X}$ rather than $\lambda_{\Sigma_X,i}$, and why all components remain represented in the distribution. This interpretation aligns with the generalized reverse water-filling picture for Gaussian vector RDP problems, where active perception constraints alter the classical single water level structure.
\end{remark}
}
\paragraph*{Continuous sources under the perfect realism regime} In what follows, we detail a new estimation method for the rate–distortion–perception function (RDPF) of multivariate continuous sources, under a single-letter average distortion constraint and within the perfect-realism regime. This approach leverages tools from copula theory, together with an information-geometric projection theorem, enabling a principled characterization of the RDPF in the perfect-realism setting \cite{serra:2024:copula}. Interestingly, the proposed framework estimates not only the RDPF under perfect realism, but also addresses two closely related companion problems: the output-constrained RDF (OC-RDF) problem \cite{saldi:2015} and the entropic optimal transport (EOT) problem \cite{wang:2022}.
\par The information theoretic definition of the OC-RDF that was formally introduced by Saldi {\it et al.} in \cite{saldi:2015} as follows.
\begin{definition}(OC-RDF)\label{problem:OC-RDF}
Let the probability density function (p.d.f.) $f_{X} \in \mathcal{P}(\mathcal{X})$. Then, the OC-RDF for the source {$X \sim f_{X}$} under a distortion measure $\Delta:\mathcal{X} \times \hat{\mathcal{X}} \to \mathbb{R}^+_0$ and a target reconstruction distribution $f_{\hat{X}} \in \mathcal{P}(\hat{\mathcal{X}})$ is given as follows:
\begin{align}
        R_{\texttt{OC}}(D) =\min_{\substack{f_{\hat{X}|X} \in \hat{\Pi}(f_{X}, f_{\hat{X}})\\ \mathbb{E}[\Delta(X,\hat{X})] \le D}}  I(X, \hat{X}) \label{opt:RDPF} 
\end{align}
where the minimization is on the convex set of Markov kernels $\hat{\Pi}(f_{X}, f_{\hat{X}}) \triangleq \{ f_{X|\hat{X}}: m_{\hat{X}}(f_{\hat{X}|X} \cdot f_{X}) = f_{\hat{X}} \}$.
\end{definition}
It should be noted that the problem of the OC-RDF specializes to the problem of RDPF subject to a single-letter distortion and the perfect realism regime, by specifying the reconstruction distribution to be equal to the source distribution (i.e., $f_{\hat{X}} = f_{X}$).
Interestingly, the OC-RDF highlights a pleasing connection to the EOT problem, whose mathematical definition is stated as follows.

\begin{definition} \cite{wang:2022} \label{problem:EOT}
{(EOT)} Let $f_{X}\in \mathcal{P}(\mathcal{X})$ and $f_{\hat{X}} \in \mathcal{P}(\hat{\mathcal{X}})$. Then, the EOT problem for $\epsilon > 0$ and distortion measure $\Delta:\mathcal{X} \times \hat{\mathcal{X}} \to \mathbb{R}^+_0$, is given by
\begin{align}
    D_{\texttt{EOT}}(\epsilon) = \min_{f_{X,\hat{X}} \in \bar{\Pi}(f_{X}, f_{\hat{X}})} \mathbb{E}\left[\Delta(X,\hat{X})\right] + \epsilon I(X,\hat{X})
\end{align}
where the minimization is over the convex set of joint p.d.f.s $\bar{\Pi}(f_{X}, f_{\hat{X}}) \triangleq \{ f_{X,\hat{X}}: m_{X}{(f_{X, \hat{X}})} = f_{X}, m_{\hat{X}}{(f_{X, \hat{X}})} = f_{\hat{X}}\}$.
\end{definition}

Notably, it can be shown that the OC-RDF and EOT problems are closely related in the sense that, for specific values of $D$ and $\epsilon$, there exists a one-to-one mapping between the sets of solutions of the two problems. In other words, we can find the solution to one problem based on the solution of the other. Although similar links have been observed for the classical RDF, they have not been extended to RDP theory. This extension was first established in \cite{serra:2024:copula}, and we state the corresponding theorem here for completeness.

\begin{theorem} \cite{serra:2024:copula} {(Connection of OC-RDF and EOT)}\label{theorem:EquivalenceEOTRD}
    Let $f_{X} \in \mathcal{P}(\mathcal{X})$ and $f_{\hat{X}} \in \mathcal{P}(\hat{\mathcal{X}})$. Then, for any $D > 0$, there exists an $\epsilon>0$ such that the OC-RDF and EOT problems are equivalent.
\end{theorem}

\par {To keep the discussion of the subsequent results self-contained, we begin by reviewing the basics of copula distributions. A more comprehensive treatment, including additional definitions and key theorems, can be found in \cite{durante:2010}.} 

\begin{definition} \cite{serra:2024:copula} {\it (Copula distribution)} For every $d \ge 2$, a $d$-dimensional copula distribution function (d.f.) is a $d$-variate d.f. on $[0,1]^d$ whose univariate marginals are uniformly distributed on $[0,1]$.
\end{definition}

\par {The next theorem, along with its two companion corollaries, shows how copulas provide a powerful framework for modeling and analyzing multivariate distributions.} 

\begin{theorem} \cite{durante:2010} ({\it Sklar's Theorem}) Let F be a $d$-dimensional d.f. with marginal d.f. $F_1, F_2, \ldots, F_d$. Let $A_j$ denote the range of $F_j$, $A_j  \triangleq F_j \left( \bar{\mathbb{R}}\right) \quad (j = 1,2, \ldots, d)$. Then, there exists a $d$-copula d.f. $C$ such that for all $(x_1,x_2, \ldots, x_d) \in \bar{\mathbb{R}}^d$,
\begin{align}
    F(x_1, \ldots, x_d) = C\left(F_1(x_1), \ldots, F_d(x_d)\right). \label{definition:copula:Sklard}
\end{align}
Such a $C$ is uniquely determined on $A_1 \times A_2 \times \cdots A_d$ and, hence, it is unique when $F_1, F_2, \ldots, F_d$ are continuous.
\end{theorem}
\begin{corollary} \cite{durante:2010} \label{corollary:SklarTheoremForPdf}
    Let $f:\bar{\mathbb{R}}^d \to \mathbb{R}_{+}$ be the probability density function (p.d.f.) associated with \eqref{definition:copula:Sklard}. Then, $f$ can be uniquely decomposed as
    \begin{align}
        f(x_1,\ldots, x_d) = c\left(F_1(x_1), \ldots, F_d(x_d)\right) \prod_{j=1}^d f_j(x_j)
    \end{align}
    where $f_j$ is the p.d.f. associated with the univariate marginal d.f. $F_j$ and $c:[0,1]^d \to \mathbb{R}_{+}$ is the p.d.f. associated with the copula d.f. $C$. 
\end{corollary}
\begin{corollary} \cite{durante:2010} \label{corollary:CopulaConstruction}
    Let $F_1, F_2, \ldots, F_d$ be univariate d.f.'s and $C$ be a copula d.f. {Then}, the function $F:\bar{\mathbb{R}}^d \to [0,1]$ defined in \eqref{definition:copula:Sklard} is a $d$-dimensional d.f. with marginal $F_1, F_2, \ldots, F_d$.
\end{corollary}
It is worth noting that Corollary \ref{corollary:SklarTheoremForPdf} guarantees that the p.d.f. of any multivariate distribution can be factorized into the product of its marginal densities and a unique copula density. This factorization can be interpreted as decoupling the dependence structure encoded in the joint distribution (captured by the copula) from the information carried by each marginal. On the other hand, Corollary \ref{corollary:CopulaConstruction} ensures that, for a fixed set of marginals, any choice of copula yields a valid joint distribution.
\par We conclude this review with the definition of the quantile function, which will also be of use in the derivation of our main results.

\begin{definition} \cite{durante:2010} \textit{(Quantile function)} Let $X \sim F_X$ be a univariate RV on $\mathcal{X} \subseteq \mathbb{R}$. We define the quantile function $Q_{X}:[0,1] \to \mathbb{R}$ as $Q_{X}(u) \triangleq \sup \{ x \in \mathcal{X}: F_X(x) \le u \}$. If $F_{X}$ is continuous and strictly increasing, then $Q_{X} = F^{-1}_X$. However, even if $F_{X}$ may fail to have an inverse function, $Q_{X}$ guarantees that $Q_{X}\left(F_{X}(X)\right) = X$ a.s.    
\end{definition}

To simplify the notation, we refer to the uniform transformation of a random vector $X = (X_1,\ldots, X_d)$ as the mapping $\Phi_{X}:\mathcal{X} \to [0,1]^d$ defined as $\Phi_{X}(X) \triangleq (F_{X_1}(X_1), \ldots, F_{X_d}(X_d))$. Moreover, we define the mapping $\Psi_{X}:[0,1]^d \to \mathcal{X}$ as $\Psi_{X}(U) \triangleq (Q_{X_1}(U_1), \ldots, Q_{X_d}(U_d))$.  By construction, $\Psi_{X}$ is the a.s.-inverse of $\Phi_{X}$, that is, $\Psi_{X}(\Phi_{X}(X)) = X$ a.s.

\par First, we state a lemma that allows the functionals appearing in the mathematical formulations of Definitions~\ref{problem:OC-RDF} and~\ref{problem:EOT} to be equivalently rewritten in terms of copula distributions.

\begin{lemma} \cite{serra:2024:copula} \label{lemma:MI2KL}
Let $(X,\hat{X}) \sim f_{X,\hat{X}} \in \mathcal{P}(\mathcal{X} \times \hat{\mathcal{X}})$ be a $2d$-variate RV with marginal p.d.f. $f_{X} \in \mathcal{P}(\mathcal{X})$ and $f_{\hat{X}} \in \mathcal{P}(\hat{\mathcal{X}})$. Then, the mutual information $I(X,\hat{X})$ can be equivalently written as follows 
\begin{align}
    I(X,\hat{X}) = \KL(C_{X,\hat{X}}||C_X \otimes C_{\hat{X}}) \label{eq:copula:generalMI}
\end{align}
where $C_{X,\hat{X}}, C_X, C_{\hat{X}}$ are the copula d.f.'s associated with distributions $F_{X,\hat{X}}$, $F_{X}$, and $F_{\hat{X}}$, respectively. In addition, given a distortion function $\Delta:\mathcal{X} \times \hat{\mathcal{X}} \to \mathbb{R}^+$, the following holds
\begin{align}
    \mathbb{E}_{F_{X,\hat{X}}}\left[\Delta(X,\hat{X})\right] = \mathbb{E}_{C_{X,\hat{X}}}[\Delta\left(\Psi_{X}(U_X),\Psi_{\hat{X}}(U_{\hat{X}})\right)] \label{eq:copula:distortion}
\end{align}
where $U = (U_X,U_{\hat{X}}) \sim C_{X,\hat{X}}$.
\end{lemma}
\par {Leveraging Lemma \ref{lemma:MI2KL}, we can provide an alternative formulation of the mathematical expression in \eqref{opt:RDPF} as follows
\begin{align}
R_{\texttt{OC}}(D) &=  \min_{\substack{C \in \mathcal{C}_{2d}\\ \mathbb{E}_C\left[{\Delta(\Psi_{X}(U_X),\Psi_{\hat{X}}(U_{\hat{X}}))}\right] = D}}  \KL(C||{C_{X} \otimes C_{\hat{X}}}) \label{opt:CopulaKL} 
\end{align}
where $\mathcal{C}_{2d}$ is the set of $2d$-copulas and $D \in [D_{\min}, D_{\max}]$.
}
\par It should be noted that \eqref{opt:CopulaKL} constitutes a convex program over the space of copula distribution functions. Additionally, this formulation admits an information-geometric interpretation as a projection problem where the objective is to determine the copula distribution $C$ that minimizes the information divergence from the independent product copula $C_X \otimes C_{\hat{X}}$, subject to a collection of linear constraints. Csisz\'ar has extensively studied such projection problems in \cite{Csiszar:1975}, where the analytical form of the optimal projection in the relevant setting is characterized. Consequently, by invoking \cite{Csiszar:1975}, we obtain the following theorem.
\begin{theorem} \cite{serra:2024:copula}{(Analytical solution of \eqref{opt:CopulaKL})} \label{theorem:OptimalProjection}
Let $R = C_{X} \otimes C_{\hat{X}} $ and assume there exists a copula d.f. $P$  such that ${\KL(P||R)} < \infty$ and the constraint in \eqref{opt:CopulaKL} is satisfied. Then, \eqref{opt:CopulaKL} admits a minimizing copula $Q$ with Radon–Nikodym derivative with respect to the measure $R$ of the form
\begin{align}
\frac{dC}{dR}(\textbf{u}) = e^{\mu + \theta[\Delta(\Psi_{X}(\textbf{u}_x),\Psi_{\hat{X}}(\textbf{u}_{\hat{X}}))]} \prod_{i = 1}^{2d} g_{i}(u_{i}) \label{eq:copula:optimaldensity}
\end{align}
for some constants $(\mu, \theta)$, and nonnegative uni-variate functions $g_i$ such that $\log(g_i(s)) \in l_1([0,1])$ for $i=1,\ldots,2d$.
\end{theorem}
Although Theorem \ref{theorem:OptimalProjection} provides a characterization of the solution of \eqref{opt:CopulaKL}, the lack of an analytical form for the free functions $\{g_i(\cdot)\}_{i=1\ldots,2d}$ poses a challenging problem in the computation of \eqref{eq:copula:optimaldensity}. To circumvent this technical issue, we consider a relaxation of the constraint set in \eqref{opt:CopulaKL}, which results in a lower bound on the OC-RDF. This is demonstrated next.
\par For any integer $N$, \eqref{opt:CopulaKL} can be lower bounded as follows
\begin{align}
R_{\texttt{OC}}(D) \ge R_{\texttt{OC}}^{(N)}(D) =\min_{\substack{Q \in \mathcal{P}([0,1]^{2d})\\ \mathbb{E}\left[{\Delta(\Psi_{X}(U_X),\Psi_{\hat{X}}(U_{\hat{X}}))}\right] = D\\
\mathbb{E}_Q[u_{i}^n] = \alpha_n,~~(i,n) \in I}}  \KL(Q|| R )\label{prob:2}
\end{align}
where $R = C_{X} \otimes C_{\hat{X}}$, $I = (1,\ldots,2d) \times (1,\ldots,N)$, $D \in [D_{\min}, D_{\max}]$, and $\alpha_n$ is the $n$-th moment of a uniform distribution on $[0,1]$. 
\noindent The main technical difference between \eqref{opt:CopulaKL} and \eqref{prob:2} lies in their constraint sets. In particular, in \eqref{opt:CopulaKL} we require the minimizing distribution $Q^\star$ to belong to the class of copula distributions, i.e., to have uniformly distributed marginals. In contrast, the minimizer $Q_N^\star$ of \eqref{prob:2} is only required to match the uniform distribution up to its first $N$ moments. This implies that the feasible set of \eqref{opt:CopulaKL} is a strict subset of the feasible set of \eqref{prob:2}, thereby justifying the latter as a lower bound.
The following theorem demonstrates that, for $N \to \infty$, \eqref{prob:2} recovers the solution of \eqref{opt:CopulaKL}.
\begin{theorem} \cite{serra:2024:copula} \label{theorem:convergenceRelaxation}
Let $Q^*$ be the optimal solution of \eqref{opt:CopulaKL} and $Q^*_N$ be the optimal solution of \eqref{prob:2}. Then, 
\begin{align*}
\KL(Q^*_N || Q^*) \xrightarrow{N\rightarrow\infty} 0  ~\text{and}~ R_{\texttt{OC}}^{(N)}(D) \xrightarrow{N\rightarrow\infty} R_{\texttt{OC}}(D).
\end{align*}
\end{theorem}
Using Theorem \ref{theorem:convergenceRelaxation}, we can clearly work out the solution of \eqref{prob:2} instead of \eqref{opt:CopulaKL}. This is stated next. 
\begin{theorem} \cite{serra:2024:copula} {(Analytical solution of \eqref{prob:2})} \label{theorem:ELProjection}
    Let $R = C_{X} \otimes C_{\hat{X}} $ and assume there exists a d.f. $P$ on $[0,1]^{2d}$ such that ${\KL(P||R)} < \infty$ and the constraint set in \eqref{opt:CopulaKL} is satisfied. Then, \eqref{prob:2} admits a minimizing distribution $Q$ with Radon–Nikodym derivative with respect to the measure $R$ of the form 
\begin{align}
\frac{dQ}{dR}(\textbf{u}) = e^{\mu + \theta\Delta(\Phi_{X}(\textbf{u}_x),\Phi_{\hat{X}}(\textbf{u}_{\hat{x}}))} \prod_{i = 1}^{2d} e^{\sum_{n=0}^N \nu_{i,n} u_{i}^n} \nonumber 
\end{align}
where the positive or negative constants $(\mu, \theta, \{ \nu_{i,n} \}_{(i,n)\in I} )$ are the Lagrange multipliers associated with \eqref{prob:2} obtained as a result of the following dual program
\begin{align}
\begin{split}
\min_{\mu, \theta, \{ \nu_{i,n} \}_{(i,n)\in I}} & \mathbb{E}_R\left[{\frac{dQ}{dR}}\right] -\mu -\theta D -\sum_{(i,n) \in I}  \nu_{i,n} \alpha_n.
\end{split}\label{eq:copula:ELdual}
\end{align}
\end{theorem}
A consequence of Theorem \ref{theorem:ELProjection}
is that the mutual information $I(X,\hat{{X}})$ with joint distribution $(X,\hat{X})$ defined by marginals d.f.s $\{F_{X_i}\}_{i = 1,\ldots,d}$ and $\{F_{\hat{X}_i}\}_{i = 1,\ldots,d}$ and copula $Q$ can be written in the following form
\begin{align}
I(X,\hat{X}) = \KL(Q||R) = -\mu -\theta D -\sum_{(i,n) \in I}  \nu_{i,n} \alpha_n \label{eq:OptimalMI}.
\end{align}

Theorem \ref{theorem:ELProjection} asserts that the Lagrange multipliers $(\mu,\theta,{\nu_{i,n}}_{(i,n)\in I})$ characterizing the optimal solution of \eqref{prob:2} can be obtained by solving the dual problem \eqref{eq:copula:ELdual}. In general, \eqref{eq:copula:ELdual} does not admit a closed-form solution; however, owing to its favorable structure, the optimal multipliers can be computed efficiently using numerical optimization methods. A key ingredient for enabling efficient numerical optimization methods is to recognize that the optimization problem \eqref{eq:copula:ELdual} is strictly convex, hence it has a unique solution.
{To compute \eqref{eq:copula:ELdual}}, we propose a low-complexity optimization scheme based on gradient methods. The main technical detail to clarify concerns the estimation of the integral present in \eqref{eq:copula:ELdual}, since numerically solving a possibly high-dimensional integral could dominate the complexity of any algorithmic embodiment. However, since its computation is required only to estimate the gradient and not to compute $I(X,\hat{X})$ (as is evident in \eqref{eq:OptimalMI}), we can approximate the integral using the Monte Carlo method. The resulting iterative scheme can be considered as a {\it mini-batch stochastic gradient descent algorithm} on a convex objective. The proposed algorithm can be found in \cite[Algorithm 1]{serra:2024:copula}.

\noindent\textbf{Analytical Expressions via Shannon Lower Bound.} In this part of the tutorial, we state an analytical expression for a lower bound on the RDPF under MSE distortion in the perfect-realism regime {which serves as a baseline for assessing the tightness of the proposed algorithm in \cite[Algorithm 1]{serra:2024:copula}}. This closed-form bound can be viewed as a generalization of the celebrated Shannon lower bound (SLB) \cite{berger:1971}, which is a cornerstone of RD theory under MSE distortion.
\begin{theorem} \cite{serra:2024:copula}\label{th:ShannonLowerbound} (SLB for PR-RDPF) Let $\mathcal{S} \triangleq \{f_{X}: \mathbb{E}_{f_X}\left[{(X - \E{X})(X -\E{X})^T}\right] \preceq \Sigma\}$ be the set of source distributions with a fixed covariance matrix $\Sigma$. Then, for all $X \sim f_{X}$ with $f_{X} \in \mathcal{S}$, the RDPF under MSE distortion constraint in the perfect realism regime, denoted as $R_{\texttt{PR}}(D)$, satisfies the following lower bound
\begin{align}
R_{\texttt{PR}}(D) \ge R^{\texttt{SLB}}_{\texttt{PR}}(D) = h(X) - h(X^*) +R^{G}_{\texttt{PR}}(D) \label{eq:shannonLB}
\end{align}
where $R^{G}_{\texttt{PR}}(D) $ denotes the Gaussian RDPF under the perfect realism regime, for a source $X^* \sim N(0, \Sigma_X)$. 
\end{theorem}
Notably, for the scalar case, the SLB in \eqref{eq:shannonLB} admits a simpler analytical expression. Specifically, by letting $\mathcal{S} \triangleq \{f_{X}: \mathbb{E}_{f_X}\left[{(X -\E{X})^2}\right] \le \sigma^2_X\}$ for a finite variance $\sigma^2_X$ we obtain \cite{serra:2024:copula}
\begin{align*}
R_{\texttt{PR}}(D) \ge R^{\texttt{SLB}}_{\texttt{PR}}(D)=\frac{1}{2} \log \left(\frac{N(X)}{D - \frac{D^2}{4\sigma_X^2}} \right)
\end{align*}
with $N(X)$ denoting the entropy power of the source $X$. For the general vector case, the SLB in \eqref{eq:shannonLB} depends on $R^{G}_{\texttt{PR}}(D)$, which can be readily computed via the adaptive reverse water-filling procedure described in Corollary \ref{th: perfect_realism}.
{
\begin{remark}(Tightness and scope of the SLB)\label{remark:tightness_slb_pr_rdpf}
The bound presented in Theorem \ref{th:ShannonLowerbound} is an implicit Shannon-type lower bound rather than an exact characterization of $R_{\texttt{PR}}(D)$. While it yields the exact perfect-realism RDPF for Gaussian sources, equation \eqref{eq:shannonLB} admits a classical SLB analytical structure only in the scalar case; vector sources exhibit non-constant distortion allocation (see Remark \ref{remark:perfect_real_water}), which prevents such an analytical and exact form. For general sources, the bound is tight only in the low-distortion regime and becomes increasingly loose at moderate and high distortions, since the entropy-maximization principles underlying the SLB are not generally tight. Therefore, $R_{\texttt{PR}}^{\texttt{SLB}}(D)$ should be used primarily as a tractable analytical benchmark, whereas numerical methods such as \cite[Algorithm]{serra:2024:copula} are needed to estimate the actual $R_{\texttt{PR}}(D)$ outside the high-resolution regime.    
\end{remark}
}

\par
\begin{figure}[ht]
\vspace{-0.8cm}
    \centering
    \subfloat[MSE]{%
        \includegraphics[width=0.47\linewidth]{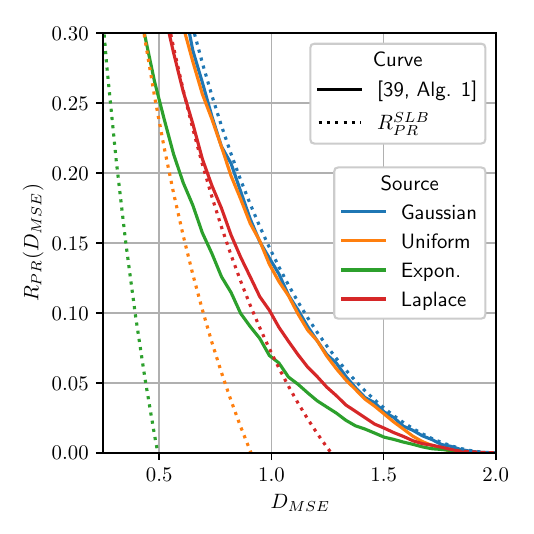}
        \label{fig:mse}
    }
    \hfill
    \subfloat[MAE]{%
        \includegraphics[width=0.47\linewidth]{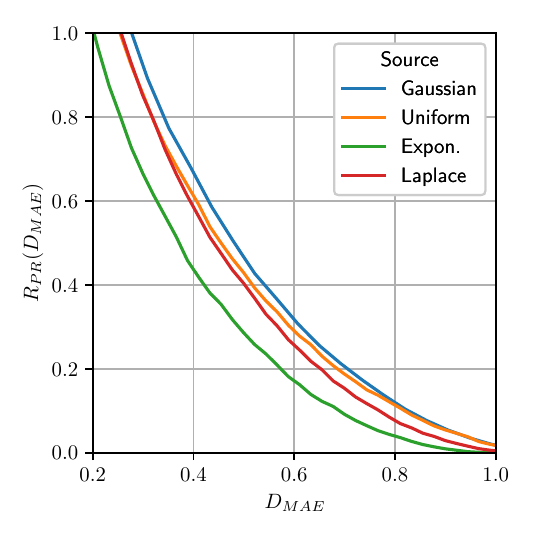}
        \label{fig:mae}
    }
\caption{$R_{\texttt{PR}}(D)$ for various source distributions under (a) the MSE distortion metric and (b) the MAE distortion metric.}
\label{fig:6}
    \vspace{-0.3cm}
\end{figure}

 In Fig. \ref{fig:6}, we illustrate $R_{\texttt{PR}}(D)$ for scalar sources under a single-letter constraint on the reconstruction error in terms of (a) the MSE distortion  (see Fig.~\ref{fig:mse}), and (b) the mean-absolute-error (MAE) distortion (see Fig.~\ref{fig:mae}). We compare the results for various source distributions, such as Gaussian, Laplace, exponential, and uniform, assuming that the source $X \sim (0,1)$, i.e., has zero mean and unit variance $\sigma_X^2 = 1$. In Fig.~\ref{fig:mse}, we also compare the estimated result with the SLB derived in Theorem \ref{th:ShannonLowerbound}. In Fig. \ref{fig:mse}, {the scalar Gaussian source provides a special case in which $R^{\texttt{SLB}}_{\texttt{PR}}(D)$ coincides with $R_{\texttt{PR}}(D)$ and therefore serves as a useful sanity check for the numerical procedure proposed in \cite{serra:2024:copula}. For the other cases {(scalar non-Gaussian sources)}, the numerical results indicate that $R^{\texttt{SLB}}_{\texttt{PR}}(D)$ behaves similarly to the SLB for the classical RDF: it is informative mainly in the low-distortion, high-resolution regime and becomes loose in the moderate- to high-distortion regimes.}

\begin{figure}[ht]
\vspace{-0.8cm}
    \centering
    \subfloat[bivariate Gaussian+MSE]{%
        \includegraphics[width=0.47\linewidth]{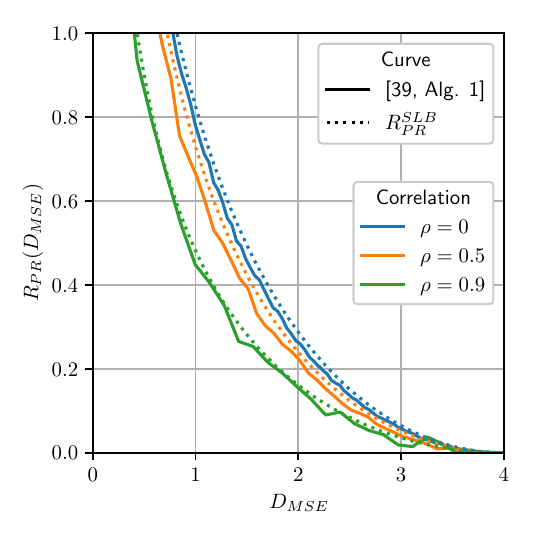}
        \label{fig:vector:gaussian}
    }
    \hfill
    \subfloat[bivariate exponential+MSE]{%
        \includegraphics[width=0.47\linewidth]{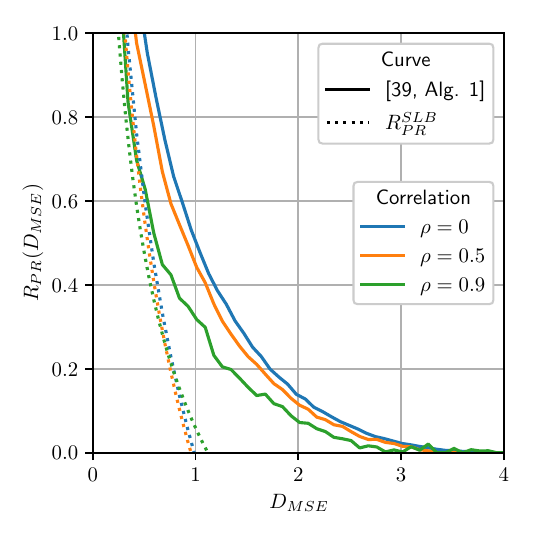}
        \label{fig:vector:exponential}
    }
\caption{$R_{\texttt{PR}}(D)$ under the MSE distortion metric for (a) a Gaussian, and (b) an exponential bivariate source.}
\label{fig:7}
    \vspace{-0.3cm}
\end{figure}

In Fig.~\ref{fig:7}, we estimate $R_{\texttt{PR}}(D)$ under an MSE distortion metric for correlated bivariate sources, considering two representative scenarios. The first involves a bivariate Gaussian source (see Fig.~\ref{fig:vector:gaussian}), while the second considers a bivariate exponential source (see Fig.~\ref{fig:vector:exponential}) with zero mean and variance $\sigma_X^2 = 1$. In both cases, the multivariate distribution is constructed by imposing a Gaussian copula with a tunable correlation coefficient $\rho \in [0,1]$ on the specified marginal distributions. By varying $\rho$, we examine settings where the marginals are independent ($\rho = 0$), mildly correlated ($\rho = 0.5$), or highly correlated ($\rho = 0.9$).
In Fig.~\ref{fig:vector:gaussian}, we compute $R_{\texttt{PR}}(D)$ using \cite[Algorithm 1]{serra:2024:copula} and compare it with the bound $R^{\texttt{SLB}}_{\texttt{PR}}(D)$ in \eqref{eq:shannonLB}, where the term $R_{\texttt{PR}}^G(D)$ is evaluated via the optimal adaptive reverse water-filling solution of Corollary \ref{th: perfect_realism}. This yields a tight $R^{\texttt{SLB}}_{\texttt{PR}}(D)$ in the Gaussian case. The numerical results show that the implementation proposed in \cite{serra:2024:copula} provides an accurate estimate of $R_{\texttt{PR}}^G(D)$ for all selected values of $\rho$. Moreover, the estimation error remains stable for low to moderate correlation, while exhibiting slightly noisier fluctuations in the high-correlation regime. In contrast, Fig.~\ref{fig:vector:exponential} shows that, beyond the high-resolution (low-distortion) regime, the $R_{\texttt{PR}}(D)$ obtained via \cite[Algorithm 1]{serra:2024:copula} is significantly tighter than $R^{\texttt{SLB}}_{\texttt{PR}}(D)$. Indeed, the latter displays behavior similar to the SLB for the classical RDF in the multivariate non-Gaussian case.

{
\subsection{Summary and Comparison with Recent Computational Frameworks}
}
\par {We conclude this section by summarizing the main computational frameworks discussed above and positioning them relative to other recent approaches for computing RDPFs. The methods reviewed in Sections \ref{section:iiia} and \ref{section:iiib} address different instances of the RDPF depending on the source model, the class of perceptual constraints, and the desired level of analytical tractability. For finite-alphabet sources with $f$-divergence perception constraints, the NAM and RAM schemes provide alternating-minimization procedures with complementary advantages: NAM enjoys broad convergence guarantees over the Lagrange-multiplier domain but requires smooth perception metrics and incurs higher per-iteration complexity, whereas RAM avoids differentiability requirements and is directly applicable to nonsmooth metrics such as total variation, at the price of potentially recovering only part of the RDP surface. For continuous sources, the Gaussian formulation leads to structured optimization problems that admit water-filling-type interpretations, while the copula-based approach provides a flexible numerical framework for estimating perfect-realism RDPFs beyond Gaussian models.}

{Recent works have developed complementary computational frameworks for discrete sources using optimal transport and primal–dual optimization. A key contribution reformulates the RDPF with Wasserstein perception constraints as a Wasserstein-barycenter problem, leveraging optimal-transport couplings and entropy regularization via alternating Sinkhorn-type methods \cite{chenc:2023}. Building on this approach, subsequent works have computed the RDPF for specific distortion and perception levels using primal-dual algorithms \cite{chenc:2025} and characterized critical transitions in RDP-related functions \cite{chenc:2024}, identifying regimes in which either the distortion or perception constraint becomes inactive. These transition phenomena are closely related to the active-constraint regimes discussed in the Gaussian setting above, where the RDP surface may reduce to the classical RDF whenever the perception constraint is inactive, or enter a genuinely perception-constrained regime when both constraints bind.
}

\begingroup
%

\begin{table*}[t!]
\centering
\small
\setlength{\tabcolsep}{6pt}
\renewcommand{\arraystretch}{1.18}
\caption{Summary of Computational Methods for RDPF. Each method is characterized by the applicable source type, constraints, key advantages, and limitations.}
\label{table:1}

{
\begingroup


\begin{tabularx}{\textwidth}{
    >{\raggedright\arraybackslash\bfseries}p{0.18\textwidth}
    >{\raggedright\arraybackslash}X
    >{\raggedright\arraybackslash}X
    >{\raggedright\arraybackslash}X
}
\toprule[1.1pt]
\textbf{Method (Ref.)} &
\textbf{Source \& Constraints} &
\textbf{Advantages} &
\textbf{Limitations} \\
\midrule

\textbf{NAM} (Th.~2) &
Finite alphabets with smooth $f$-divergence &
Exponential convergence; valid for any $(s_D, s_P)$; no multiplier restrictions &
Requires $\mathcal{C}^2$ divergence; high per-iteration cost \\

\textbf{RAM} (Th.~3) &
Finite alphabets with $f$-divergence &
No differentiability required; lower cost; handles TV; directly implementable &
Restricts the $s_P$ range; may not recover the full curve \\

\textbf{Gauss-Seidel} (Th.~6) &
Multivariate Gaussian with tensorizable constraints &
Vector sources; systematic allocation; perfect realism (Corollary~1); waterfilling interpretation &
Computationally intensive for large $d$; requires tensorizability; sublinear convergence \\

\textbf{Copula} (Th.~10-12) &
Multivariate, Continuous (non-Gaussian) \newline Perfect realism &
Non-Gaussian sources; few limitations on the distortion metric &
No closed-form expressions for $g_i$'s; numerical approximation via Monte Carlo integration \\

\textbf{Wasserstein barycenter / Sinkhorn} \cite{chenc:2024} 
& Finite alphabets with Wasserstein, KL, or TV perception constraints 
& Natural for Wasserstein perception; uses an alternating Sinkhorn implementation;
& Depends on regularization and transport dimension; convergence requires fine-tuning of additional parameters \\

\textbf{Primal--dual algorithm} \cite{chenc:2025} 
& Finite alphabets with relaxed reconstruction-distribution optimization 
& Provable convergence; explicit convergence-rate guarantees; general convex-optimization viewpoint 
& Developed for finite alphabets; practical performance depends on constraint structure \\

\bottomrule[1.1pt]
\end{tabularx}

\arrayrulecolor{black}
\endgroup
}
\end{table*}



\begin{table*}[t]
\centering
\small
\setlength{\tabcolsep}{6pt}
\renewcommand{\arraystretch}{1.18}

\caption{Computational complexity and convergence comparison. When available, total complexity for achieving $\epsilon$-accuracy is stated; otherwise, per-iteration complexity and iteration count parameters are reported.}
\label{table:2}

{
\begingroup


\begin{tabularx}{\textwidth}{
    >{\raggedright\arraybackslash\bfseries}p{0.17\textwidth}
    >{\raggedright\arraybackslash}X
    >{\raggedright\arraybackslash}X
    >{\raggedright\arraybackslash}X
}
\toprule[1.1pt]
\textbf{Method} &
\textbf{Per-Iteration Cost} &
\textbf{Convergence Rate} &
\textbf{Total Complexity} \\
\midrule

NAM &
$O(|\hat{\mathcal{X}}|^3)$ \emph{(Jacobian inversion)} &
$O(\log(1/\epsilon))$ &
$O(|\hat{\mathcal{X}}|^3 \log(1/\epsilon))$ \\

RAM &
$O(1)$ &
$O(\log(1/\epsilon))$ &
$O(\log(1/\epsilon))$ \\

Gauss--Seidel &
$O(1/\sqrt{\epsilon})$ \emph{(sol. subproblems)} &
$O(1/\epsilon^2)$ \emph{(upper bound)} &
$O(1/\epsilon^{2.5})$ \emph{(upper bound)} \\

Copula &
$O(1/\epsilon^2)$ \emph{(Monte Carlo estimation)} &
$O(1/\epsilon^2)$ \emph{(SGD)} &
$O(1/\epsilon^4)$ \\

Wasserstein barycenter / improved AS \cite{chenc:2024}
& $O\!\left(T_{\rm AS}N_{\rm Newt}
\left(|\mathcal X||\hat{\mathcal X}|+|\hat{\mathcal X}|^2\right)\right)$ 
& Empirical convergence for the entropy-regularized problem; convergence to the original WBM-RDP as $\epsilon\to 0$. 
& No explicit $\epsilon$-accuracy iteration complexity is provided in \cite{chenc:2024}; $T_{\rm AS}$ denotes the number of outer AS iterations, and $N_{\rm Newt}$ denotes the number of Newton steps.\\

Primal--dual algorithm \cite{chenc:2025}
& First-order primal--dual updates 
& $O(1/n)$ on average \textit{(lower bound)}
& Problem-dependent per-iteration cost \\

\bottomrule[1.1pt]
\end{tabularx}

\arrayrulecolor{black}
\endgroup
}
\end{table*}

%
%
%
%
%
\arrayrulecolor{black}
\endgroup

{Broadly speaking, Wasserstein-barycenter methods are especially natural when the perceptual fidelity measure is induced by optimal transport; NAM and RAM are tailored to finite-alphabet $f$-divergence constraints; Gaussian methods exploit continuous-source structure to obtain analytical or semi-analytical characterizations; and copula-based methods target non-Gaussian continuous sources in the perfect-realism regime. These approaches should therefore be viewed as complementary rather than competing, each covering a different region of the broader computational landscape of RDP theory.
}

\par {Tables~\ref{table:1} and~\ref{table:2} summarize the scope, advantages, limitations, and computational complexity of these representative methods.
}

\section{Promising Future Research Directions}\label{sec:future_work}
In this section, we discuss a few promising research directions that naturally emerge from this line of work. A first and fairly direct extension is to bring RDP theory into the realm of multi-user information theory \cite{gamal:2011}, a direction that has already begun to attract attention in the literature. In this tutorial, however, we focus on a few additional research avenues that are particularly intriguing, as they lie at the intersection of machine learning and information-theoretic control.

\subsection{Neural compression for perception-aware lossy compression}\label{subsec:neural_compression}

A strong future research direction, emerging in part from the optimization perspectives of this tutorial, is to move beyond classical RD optimization and treat RDP coding as a first-class design principle for modern neural compression. The central idea is simple: compression systems should not only minimize distortion at a given bitrate, but should also explicitly control perceptual fidelity. In other words, reconstructions should look natural and match the source distribution, even when pixel-wise error is not driven to zero. What makes this direction especially exciting from a coding-theoretic perspective is that, as shown in \cite{lei:2025}, structured coding tools such as lattice quantization can be combined with neural compressors to obtain strong RD efficiency through geometric packing. At the same time, perceptual constraints may require an additional degree of freedom. This can be realized as shared randomness between encoder and decoder, implemented constructively via shared dithering across lattice cells. This view opens a new design space: Neural estimators, structured quantizers, and controlled randomness can jointly determine the operating point on the RDP frontier. This suggests a clear roadmap toward practical, perception-aware codecs that remain both theoretically grounded and empirically competitive.

 \begin{figure}[ht]
    \centering 
    \includegraphics[width=0.47\textwidth]{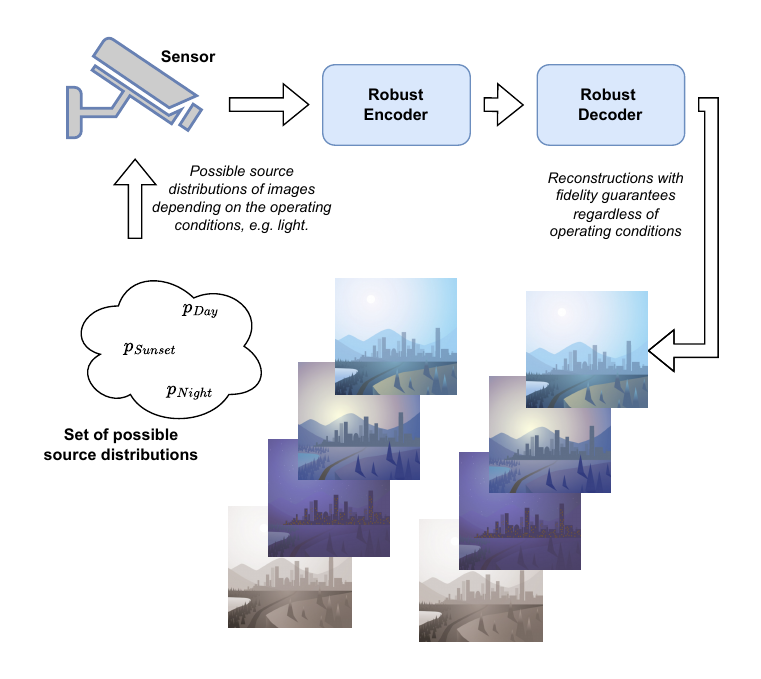}
    \caption{A distributionally robust scenario in image compression.}\label{fig:8}
\end{figure}

\subsection{Distributionally robust perception-aware lossy compression}\label{subsec:future:1}
The precursors of RDP theory \cite{blau:2019}, as well as the vast majority of related works to date, typically assume a source with known statistics, \textit{mutatis mutandis} with classical rate--distortion (RD) theory. In practice, however, the source statistics are rarely available, and one may instead seek to identify patterns in the observed data to compress unknown sources efficiently on average and asymptotically, \`a la the Lempel--Ziv algorithm \cite{ziv:1977} adapted to the context of lossy compression \cite{yang:1996} with a perceptual constraint. That said, in many mission-critical applications, such as spacecraft control and telemetry links, avionics flight-control systems, and medical devices, universal compression schemes like Lempel--Ziv may be undesirable, as they can yield variable-length representations and unpredictable delays depending on the realized data sequence. In such settings, reliability and strict timing constraints often outweigh average compression gains, making worst-case encoder/decoder design more appropriate to guarantee bounded packet sizes, deterministic latency, and safe operation even under abnormal conditions.
\par The problem of designing the worst-case encoder/decoder falls under the realm of distributionally robust lossy source coding \cite{sakrison:1969}, which addresses the fundamental challenge of performing source coding when the source statistics are uncertain. In particular, the true source distribution is \textit{unknown} but is assumed to lie within a prescribed uncertainty set of candidate distributions. A simple motivating example related to image compression is shown in Fig. \ref{fig:8}, where images taken under different lighting conditions have different distributions and a sensor is able to compress them with the same quality. Nonetheless, the scope of this framework extends far beyond that illustration. Indeed, such uncertainty naturally arises in many real-world settings. For example, in the compression of IoT sensor measurements under time-varying environmental conditions, or in the design of distributionally resilient deep neural network–based codecs, where mismatches in the source statistics can significantly affect compression performance. A preliminary work on this research direction can be found in \cite{serra:2025itw}.

\subsection{Rate-cost-perception tradeoffs in control}\label{subsec:controlled_networks}
 
\begin{figure}[ht]
    \centering 
\includegraphics[width=\columnwidth,height=4.5 cm]{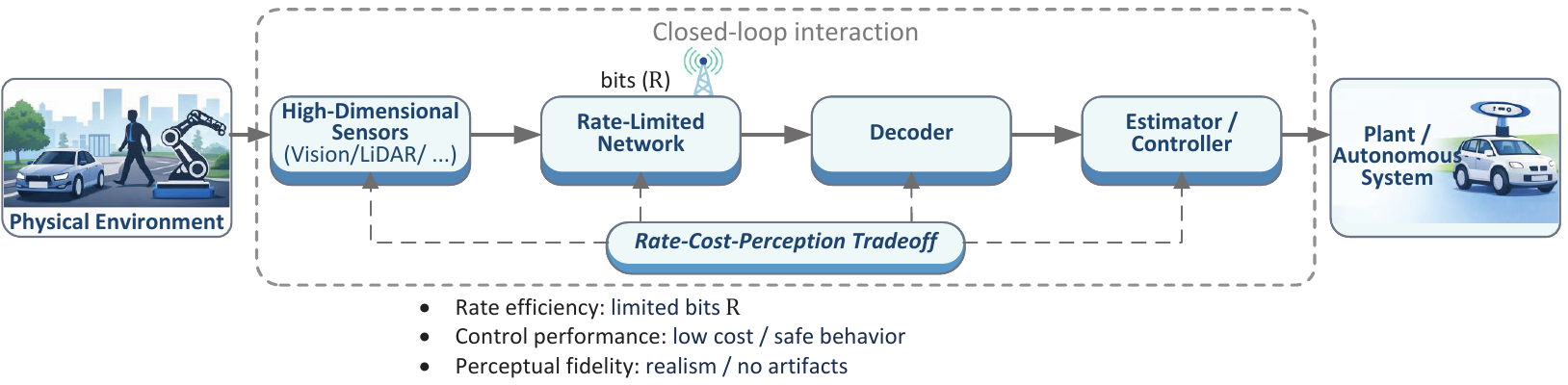}
    \caption{Perception-aware closed-loop control.}\label{fig:9}
\end{figure}

Another promising research direction is to move beyond the classical “bit-pipe” abstraction of communication constraints and develop a unified view of rate–distortion–perception tradeoffs in networked control systems \cite{hespanha:2007}. While information-theoretic control has traditionally characterized fundamental limits in terms of stabilization rates, delay/reliability requirements, and feedback capacity, future autonomous systems will increasingly be shaped by high-dimensional perception streams (vision, LiDAR, multimodal sensing) whose compression cannot be judged solely by metrics such as MSE or relevant quadratic costs. Instead, controllers and estimators will operate on reconstructions that must be simultaneously rate-efficient, task-sufficient, and perceptually consistent, where perception constraints enforce statistical realism and prevent unsafe artifacts or hallucinations that could mislead downstream modules. This opens the door to new ``control-aware'' notions of distortion and perception, new fundamental limits on the achievable tradeoff between communication rate, perceptual fidelity, and control cost, and new co-design methodologies where encoding, inference, and control policies are optimized jointly. Such a perspective is particularly relevant for safety-critical networked autonomy, like, for instance, remote driving, robot teleoperation, and industrial edge control, where the right representation is not necessarily the one that best reconstructs signals, but the one that best preserves the information that matters for reliable and safe closed-loop behavior. An illustration of the aforementioned discussion is shown in Fig. \ref{fig:9}.

\section{Conclusion}\label{sec:conclusion}
In this tutorial, we mainly focused on an anthology of generic optimization and computational tools for evaluating the first characterization of the RDPF introduced in \cite{blau:2019} for general $\IID$ source models and perceptual constraints. Along the way, we reviewed the main coding theorems that lead to different characterizations of the RDPF, highlighting the role of perception constraints in shaping the fundamental limits of lossy compression. Finally, motivated by our ongoing work in this area, we outlined a few promising research directions at the intersection of machine learning and information-theoretic control that point to new challenges and opportunities for future research.

{
\section*{Acknowledgment}
}
{This work has received funding from the European
Research Council (ERC) under the European Union’s Horizon 2020 Research and Innovation Programme (Grant agreement No. 101003431),  from the SNS JU project 6G-GOALS under the
EU’s Horizon programme (Grant Agreement No. 101139232) and the Huawei France-EURECOM Chair on Future Wireless Networks.
}

\bibliographystyle{IEEEtran}
\bibliography{string,biblio}

\end{document}